\documentclass[10pt, journal, letterpaper]{IEEEtran}
\usepackage[style=ieee,maxnames=4,minnames=3,maxbibnames=3]{biblatex}

\usepackage{amsfonts}
\usepackage[cmex10]{amsmath}
\usepackage{amsthm}

\newcommand{\mr}{\mathrm}
\def\d{\mathrm{d}}
\theoremstyle{definition}
\newtheorem{definition}{Definition}

\theoremstyle{remark}

\newtheorem{theorem}{Theorem}
\theoremstyle{remark}

\usepackage{array}
\usepackage{mdwmath}
\usepackage{mathrsfs}
\usepackage{mdwtab}
\usepackage{xurl}
\usepackage{eqparbox}
\usepackage[caption=1]{caption}
\usepackage{stfloats}
\usepackage{url}
\usepackage{amssymb}
\usepackage{float}
\usepackage{indentfirst}
\usepackage{makeidx}
\usepackage{tabularx}
\usepackage{cases}
\usepackage{algorithmic}
\usepackage{mathtools}
\usepackage{color}
\usepackage{bm}
\usepackage{stfloats}
\usepackage{lipsum}
\usepackage{amsfonts}
\usepackage{bm}
\usepackage{dsfont}
\usepackage{lipsum}
\usepackage{graphicx}
\usepackage{url}
\usepackage{bbm}
\newcommand{\RNum}[1]{\uppercase\expandafter{\romannumeral #1\relax}}
\ifCLASSOPTIONcompsoc
\usepackage[caption=false, font=normalsize, labelfont=sf, textfont=sf]{subfig}
\else
\usepackage[caption=false, font=footnotesize]{subfig}
\fi
\usepackage{multirow}
\allowdisplaybreaks
\makeatletter

\usepackage[linesnumbered,ruled,vlined]{algorithm2e}
\usepackage{amsmath}
\makeatother

\usepackage{amsthm}
\theoremstyle{definition} 
\theoremstyle{remark} 
\begin{document}
\title{
Non-Identical Diffusion Models in MIMO-OFDM Channel Generation
}

\author{Yuzhi~Yang, Omar~Alhussein, Weijie Zhou, Zhaoyang Zhang, and M\'erouane~Debbah\\
\thanks{Y. Yang, O. Alhussein, and M. Debbah are with College of Computing and Mathematical Sciences, Khalifa University, Abu Dhabi 127788, UAE (e-mails: \{yuzhi.yang, omar.alhussein, merouane.debbah\}@ku.ac.ae). }
\thanks{W. Zhou and Z. Zhang are with 1) College of Information Science and Electronic Engineering, Zhejiang University, Hangzhou 310027, China, and 2) Zhejiang Provincial Key Laboratory of Info. Proc., Commun. \& Netw. (IPCAN), Hangzhou 310027, China (e-mails: \{wj\_zhou, zhzy\}@zju.edu.cn). }}

\maketitle
\begin{abstract}
We propose a novel diffusion model, termed the non-identical diffusion model, and investigate its application to wireless orthogonal frequency division multiplexing (OFDM) channel generation.
Unlike the standard diffusion model that uses a scalar-valued time index to represent the global noise level, we extend this notion to an element-wise time indicator to capture local error variations more accurately.
Non-identical diffusion enables us to characterize the reliability of each element (e.g., subcarriers in OFDM) within the noisy input, leading to improved generation results when the initialization is biased. Specifically, we focus on the recovery of wireless multi-input multi-output (MIMO) OFDM channel matrices, where the initial channel estimates exhibit highly uneven reliability across elements due to the pilot scheme.
Conventional time embeddings, which assume uniform noise progression, fail to capture such variability across pilot schemes and noise levels.
We introduce a matrix-valued time indicator that matches the input size to control element-wise noise progression. Following a diffusion procedure similar to existing methods, we formulate the proposed non-identical diffusion scheme and evaluate it numerically. For MIMO-OFDM channel generation, we propose a dimension-wise time embedding strategy. We also develop and compare multiple training and generation methods through numerical experiments.
\end{abstract}

\section{Introduction}
\subsection{Motivation}

Diffusion models have recently emerged as a powerful class of generative models and have achieved remarkable success in a wide range of generation tasks \cite{ddpm, ddim}. In a typical diffusion process, Gaussian white noise is gradually added to a clean target, and generation is performed by approximately inverting this process, progressively removing noise. While sampling from pure noise often yields high-quality samples that follow the dataset distribution, it provides little control over the generation direction. In many practical scenarios, however, we are interested in refining a coarse or imperfect estimate toward a specific target output rather than generating from scratch.

\begin{figure}
    \centering
    \includegraphics[width=\linewidth]{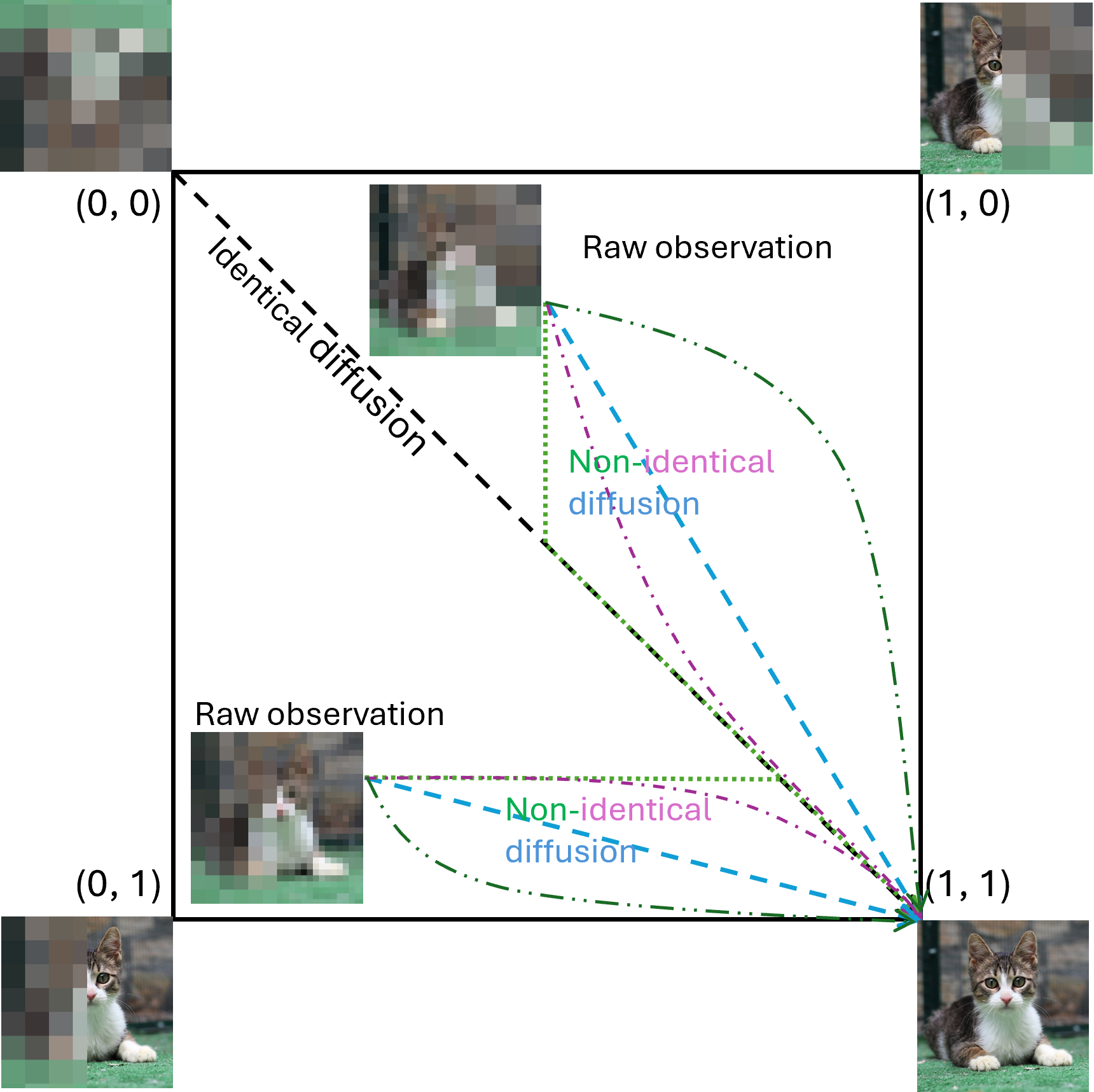}
    \caption{Illustration of identical and non-identical diffusion for MIMO-OFDM channel refinement under non-uniform element-wise reliability.}
    \label{fig:overlook}
\end{figure}

Extending this perspective to wireless transceivers, similar problems of generation from rough initialization are widespread. Unlike traditional neural network (NN)-based mapping frameworks, generative methods are gaining increasing attention due to their resilience to ubiquitous noise in wireless systems \cite{8054694, 11017513, zhu2025wirelesslargeaimodel}.
However, existing initialization strategies implicitly assume that all elements of the diffusion variable share the same reliability, which is not accurate in many applications, such as wireless multi-antenna OFDM receivers. In practice, certain elements, especially those related to pilots, are much more reliable than others, leading to a highly non-uniform noise distribution across elements. As illustrated in Fig.~\ref{fig:overlook}, a conventional diffusion model uses a single scalar timestep and therefore assumes a uniform noise level across all entries of the channel matrix. In contrast, practical MIMO-OFDM receivers often produce coarse channel estimates with highly non-uniform reliability across pilots, antennas, and subcarriers. The proposed non-identical diffusion uses an element-wise time indicator to describe this heterogeneous reliability and to guide denoising from such structured initializations.

In standard diffusion models, the noise level is encoded by a scalar time variable and it is embedded onto the size of the input. This scalar representation is fundamentally mismatched with scenarios where different entries exhibit different noise levels. Therefore, it is more desirable to represent element-wise reliabilities through a time vector or matrix, enabling a more precise description of the diffusion process. In this work, we investigate diffusion models with non-identical time for each element, referred to as \emph{non-identical diffusion}, and demonstrate their application to the MIMO-OFDM channel generation problem, which always has non-identical initialization. We focus on channel recovery from imperfect estimates that exhibit non-identical noise distributions across elements and design our approach to remain compatible with NN architectures proposed in prior diffusion-based works.

A central challenge in non-identical diffusion lies in designing an appropriate embedding mechanism. Since the original time matrix already matches the size of the input, conventional scalar time embeddings are no longer directly applicable. Exploiting the fact that resources in MIMO-OFDM systems are typically organized along time slots, antennas, and subcarriers, and building on existing dimension-wise NN architectures for channel-related tasks \cite{mixer}, we propose a column- and row-wise time embedding scheme tailored to non-identical diffusion. We further investigate how the training strategy, diffusion trajectory, and initialization distribution affect performance under this new framework.

\subsection{Related Works}

Diffusion models were first introduced for image generation and have rapidly transcended this original scope to become a widely used class of generative models across numerous domains \cite{ddpm, ddim, xu2023versatile, qiu2023irdm, chi2024rf, fesl_diffusion-based_2024, MIMOdiffusion}. Recent overviews and surveys \cite{yang2023diffusion, chen2024overview} emphasize this extensive versatility, highlighting how their denoising process allows for superior performance in various vision tasks. Specifically, diffusion models have set new benchmarks in image restoration tasks such as denoising \cite{ddpm, ddim} by effectively reversing corruption processes, and they have demonstrated exceptional capability in tasks like inpainting and completion \cite{lugmayr2022repaint, corneanu2024latentpaint, liu2023image}, where they synthesize coherent content for missing regions. Moreover, their robust semantic understanding has facilitated advanced image editing applications \cite{kawar2023imagic, huang2025diffusion, zhang2023inversion}, proving their adaptability and broad impact in the field of computer vision.

In wireless systems, generative methods have also been investigated as robust alternatives to deterministic NN mappings, particularly due to their ability to model uncertainty and mitigate measurement noise \cite{8054694, 11017513, zhu2025wirelesslargeaimodel}. \cite{sortino2024radiff, luo2024rm, wang2024radiodiff} utilize DMs to construct a radio map, with the goal of generating spatial signal strength maps from sparse samples or from past estimations. In MIMO-OFDM wireless communication systems, the work in \cite{MIMOdiffusion, zhou2025generative, chen2025generative} models channel estimation as an inverse problem, using the received signal and pilot sequence as conditional information to derive a score function. The function is then introduced into the posterior sampling formula to guide the model in generating the target result. The work in \cite{zilberstein_joint_2024, tnse} use DMs for joint channel estimation and symbol detection, to incorporate both the discrete prior of the symbols and a learned prior for the channel. \cite{yang2025diffusion, lee2025generating} introduces conditional information such as positions and coarse estimates into the denoising network, allowing the denoising network itself to estimate the noise based on these conditions. These works demonstrate that diffusion-based generative models can effectively capture complex structures of the channel state information (CSI) and uncertainties in wireless environments.

However, in the settings of tasks such as image restoration, channel estimation, and symbol detection, we often have access to noisy or low-resolution coarse observations, with the expectation of generating high-quality results based on them. Since the inference stage of DMs is essentially a denoising process, starting generation directly from Gaussian noise is not always necessary. Considering that parts of the ground truth are known during image inpainting, works \cite{lugmayr2022repaint, yu2023freedom} deviate from the standard DDPM model. At each inference step, they replace the corresponding parts of the generated result with the known observations and perform resampling to achieve semantic alignment. In handling low-resolution image restoration, \cite{wang2023dr2} avoids this direct replacement. Instead, it uses a low-pass filter to replace the low-frequency components of the generated value at an intermediate timestep with the low-frequency components of the observations. Additionally, to reduce the number of sampling steps, the generation process begins from coarse observations, and the corresponding starting timestep is also estimated. Similarly, in communication tasks, work by \cite{fesl_diffusion-based_2024} has shown that initializing the diffusion process with imperfect MIMO channel estimates can generate higher-quality channels. By modeling the relationship between the noise variance of a noisy observation and the timestep, works \cite{wang2025erasing, mohsin2025conditional} use the noisy observation directly as the starting point of the generation process, achieving progressive denoising. These approaches do not require additional network training. Simply using the coarse observations as the starting point for inference naturally enables DMs to generate the expected and semantically aligned results based on them. In contrast, if inference were to start from Gaussian noise, we would need to employ methods such as inverse problem modeling \cite{darassurvey, zilberstein2024solving} to allow the coarse observations to guide the model's generation.

 We usually need to initialize the current time in the diffusion procedure based on the predicted distance between the initial value and the target if starting from the coarse observations. However, the aforementioned works adopt a diffusion formulation where all elements of the state share the same time step and thus the same noise level, implicitly treating their reliabilities as identical. This simplifying assumption is common in diffusion-based methods \cite{ma_diffusion_2024, tnse, lugmayr2022repaint, corneanu2024latentpaint, liu2023image, kawar2023imagic, huang2025diffusion}, even when different regions of the input have very different noise characteristics. Work~\cite{mixer} proposes a dimension-wise neural architecture that exploits the inherent structure of the CSI across subcarriers and antennas through an MLP-based design. This method offers valuable insight for embedding timesteps into CSI at an element-wise level across different dimensions.

\subsection{Contributions}
In this paper, we propose and discuss a new kind of diffusion model, namely the non-identical diffusion model. Unlike traditional diffusion models, we use a matrix time indicator with the same shape as the input to better characterize the current error during the diffusion process. This structure provides a new method of generating from a rough estimation with an unbalanced noise distribution, which is common in wireless OFDM receivers. Through numerical results, we show the effectiveness of the proposed non-identical diffusion model, and analyze the impact of the noise-adding scheme during training, the initialization pattern, and the time propagation strategy during generation.
The contributions of this paper are summarized as follows:

\begin{itemize}
\item We propose the non-identical diffusion model, formally define its structure, and present its forward and denoising formulations together with its relationship to standard diffusion models.
\item We propose an interleaved time embedding method tailored for non-identical diffusion. When applied to the multi-antenna OFDM channel generation task based on multi-layer perceptron (MLP)-Mixer backbone, the proposed scheme outperforms the traditional diffusion scheme.
\item We provide numerical evaluations to validate the proposed framework and analyze the effects of the noise injection strategy during training, the initialization pattern, and the time propagation scheme during generation.
\end{itemize}

\subsection{Notations}
In this paper, non-bold symbols $x$ denote scalars or functions whose outputs are scalars. Bold italic lowercase letters $\bm{x}$ represent vectors or functions whose outputs are vectors. Bold italic uppercase letters $\bm{X}$ denote vector-valued random variables corresponding to the respective lowercase letters. Bold uppercase letters $\mathbf{X}$ denote matrices. The operator $\circ$ denotes the Hadamard product, and all scalar functions applied to vectors are understood as element-wise operations.
Without loss of generality, although diffusion variables are complex-valued matrices in the application, they are represented as real vectors of length $d$ for algorithmic presentation.
Additionally, $\|\cdot\|_2$ denotes the $\ell$-2 norm of a vector. 
$\mathcal{R}$ and $\mathcal{C}$ denote the sets of real and complex numbers, respectively. $\mathbb{E}$ denotes expectation. 
$\bm{0}$ and $\bm{1}$ represent all-zero and all-one vectors, respectively, and $\mathbb{I}$ denotes the identity matrix.

\subsection{Paper Organization}
This paper is organized as follows. Section I provides a comprehensive introduction including the motivation, a brief description of the idea, and the related works. In Section II, we briefly introduce the wireless MIMO-OFDM channel estimation problem and provide a general problem formulation. The proposed non-identical diffusion model is introduced in Section III and IV, which are based on a general generation problem. Particularly, Section III describes the proposed method from a theoretical aspect, and Section IV converts the results into operable NN solutions. Further, Section V goes back to the specific MIMO-OFDM channel estimation problem, and tackles the details in NN structure design. Section VI provides numerical results, verifying the proposed non-identical diffusion scheme. Finally, Section VII concludes the paper and provides some future directions.
\section{Problem Formulation}
In this paper, we consider a typical OFDM setting with a multi-antenna BS and a single-antenna UE. Thus, the channel can be denoted as $\mathbf{H}\in\mathcal{C}^{N_{\mr a}\times N_{\mr c}}$, where $N_{\mr a}$ and $N_{\mr c}$ denote the numbers of BS antennas and subcarriers, respectively. This model can also be regarded as a slicing of a MIMO-OFDM system corresponding to each UE.

Typically, we have some pilots distributed over several subcarriers and antennas for downlink transmission. For the channel elements not corresponding to any pilot, we cannot provide any estimation without utilizing the channel prior knowledge. For those corresponding to pilot symbols, they are always incorporated with noise. Although in some simple cases, the expected error power of each estimated channel element is the same, they may differ in many instances. For instance, when there are narrow-band interference sources that only affect several subcarriers, the estimation quality of the affected subcarriers becomes significantly worse than that of the others. Meanwhile, when we apply MIMO systems, we can use the information from other data streams to assist the estimation. In practical MIMO-OFDM systems, the coarse channel estimate usually exhibits non-uniform element-wise error variances due to the pilot allocation, interpolation strategy, and instantaneous noise realization. Representative pilot-aided OFDM/MIMO-OFDM channel estimation methods include block- and comb-type pilot arrangements with interpolation, low-rank/LMMSE refinement, and pilot-tone placement or multi-symbol training designs for reducing estimation overhead \cite{Coleri2002,Barhumi2003}.

In this paper, we do not focus on the detailed implementation of the coarse channel estimator. Instead, we assume that a coarse estimate $\overline{\mathbf H}$ and its element-wise reliability information are already available and satisfy
\begin{equation}
\overline{\mathbf H} = \mathbf{M}\circ\mathbf{H} + \mathbf{P} \circ \mathbf{N},
\end{equation}
where $\mathbf{M} \in \{0,1\}^{N_{\mr a}\times N_{\mr c}}$ indicates a mask determined by the pilot scheme, the matrix $\mathbf{P}\in \mathcal{R}_+^{N_{\mr a}\times N_{\mr c}}$ denotes the element-wise standard deviation of the estimation error, and $\mathbf{N} \sim \mathcal{CN}(\mathbf{0}, \mathbb{I})$ is standard complex Gaussian noise. In practical systems, $\mathbf{M}$ is determined by the pilot pattern, while $\mathbf{P}$ can be obtained from the adopted coarse estimator or from an analytical/empirical error-variance model. For example, LMMSE-type estimators admit explicit MMSE/error-covariance expressions \cite{Savaux2013}, while Bayesian/message-passing estimators such as sparse Bayesian learning and hybrid message passing provide approximate posterior beliefs from which element-wise uncertainty measures can be extracted \cite{Pedersen2012,Liu2023SHMP}. As in \cite{MIMOdiffusion, yang2025diffusion}, we do not explicitly formulate the channel estimation procedure, and take the raw estimation result as our input.

Although the case $\mathbf{M}=\mathbf{1}$ is not meaningful as a pilot-aided estimator, we still include it as a synthetic reference setting for methodological comparison. We also note that, as we assume the estimation errors of different channel matrix elements are independent, $\mathbf{M}$ and $\mathbf{P}$ can be represented by one matrix through simple element-wise amplification. Specifically, we can write it in the following form, which is closer to the expression in traditional diffusion models.
\begin{equation}\label{init}
\widetilde{\mathbf H} = \mathbf{A}\circ\mathbf{H} + \mathbf{B} \circ \mathbf{N}, \quad\mathbf{A}^2+\mathbf{B}^2=\mathbf{1},
\end{equation}
which requires a simple mapping that
\begin{equation}\begin{aligned}
\widetilde{H}_{i,j}={\bar{H}_{i,j}}/{\sqrt{M^2_{i,j}+P_{i,j}^2}},\\
A_{i,j}=M_{i,j}/{\sqrt{M^2_{i,j}+P_{i,j}^2}},\\
B_{i,j} ={P_{i,j}}/{\sqrt{M^2_{i,j}+P_{i,j}^2}}.
\end{aligned}
\end{equation}
In the following parts of this paper, we consider a simplified problem: Given a rough initial channel estimation as in \eqref{init}, how to generate high-quality full channel that is close to the ground truth?

\section{Non-Identical Diffusion Process}
In this section, we provide the definition of the proposed non-identical diffusion process. Throughout this paper, we refer to the traditional diffusion model as ``identical diffusion model.'' Although the diffusion variable is usually a matrix or tensor, we use its vectorized form when discussing the diffusion process. The idea is simple that we substitute the scalar time indicator $t$ in identical diffusion models by a vector $\bm{t}$ with the same size as the diffusion variable. Since there is a unitary mapping between the time indicator $t$ and the noise level in diffusion models, we can directly extend it to a vector form to represent the non-identical noise.
\subsection{Noise adding process}
Similar to the identical diffusion process \cite{ddpm}, we define the non-identical diffusion process as follows:
\begin{definition}\label{def:1}
A random process involving variable $\bm{h}\in\mathcal{R}^d$ and time $\bm{t}\in[0, T)$ is called a non-identical diffusion process if it satisfies all the following conditions:
\begin{enumerate}
    \item $\bm{\alpha}_t: [0, T)\rightarrow (0,1]^d$ is a derivable non-increasing function such that $\bm{\alpha}_{t_1}\preceq\bm{\alpha}_{t_2}$ for any $t_1>t_2$. 
    \item $\bm{\alpha}_0=\bm{1}$ and $\lim_{t\rightarrow T}\bm{\alpha}_t=\bm{0}$.
    \item $\bm{H}_{0} \sim \mu$, which is the distribution of the dataset.
    \item With $\bm{W}_t$ representing the $d$-dimensional standard Brownian motion, the following It\^o stochastic differential equation (SDE) holds:
    \begin{equation}
        \d \bm{H}_t = [\log\bm{\alpha}_t]'\circ\bm{H}_t \d t + \sqrt{-2[\log\bm{\alpha}_t]'}\circ\d \bm{W}_t.\label{SDE_forward}
    \end{equation}
\end{enumerate}
\end{definition}

Unlike the identical diffusion process, Definition \ref{def:1} substitutes the scalar $\alpha_t$ with a vector. With a well-designed path $\bm{\alpha}_t$, it can describe any monotonic path from (1,1) to (0,0) as shown in Fig. \ref{fig:overlook}. Thus, we can always design a non-identical diffusion process under Definition \ref{def:1} that passes through an arbitrarily given state following \eqref{init}.

From Definition \ref{def:1}, we can prove the following probability measure flow, which is a trivial expansion of identical diffusion.
\begin{theorem}\label{the:forward}
The law of $\bm{H}_t$ in Definition \ref{def:1} is the same as that of $\bm{\alpha}_t\circ\bm{H}_0+\bm{\beta}_t\circ\bm{\xi}$, where $\bm{\beta}_t:=\sqrt{\bm{1}-\bm{\alpha}^2_t}$, and $\bm{\xi}\sim \mathcal{N}(\bm{0}, \mathbb{I})$ is independent of $\bm{H}_0$, i.e., the distribution of $\bm{H}_t$ is
\begin{equation}
\phi^{\bm{\alpha}}_t(\bm{h}) = \int_{\mathcal{R}^d}\rho_{\bm{\beta}_t}(\bm{h}-\bm{\alpha}_t\circ{\bm{\varepsilon}})\mu(\d \bm{\varepsilon}),\label{SDE_eq}
\end{equation}
where $\rho_{\bm{\beta}}$ denotes the PDF of Gaussian distribution $\mathcal{N}\left(\bm{0},\textrm{diag}(\bm{\beta}^2)\right)$.
\end{theorem}
\begin{proof}
See Appendix \ref{app:proof1}.
\end{proof}

We also provide an intuitive understanding of this theorem. With a small positive $\Delta t$, we can approximate \eqref{SDE_forward} as
\begin{equation}
\bm{H}_t-\bm{H}_{t-\Delta t} = \frac{\bm{\alpha}_t-\bm{\alpha}_{t-\Delta t}}{\bm{\alpha}_{t-\Delta t}}\circ \bm{H}_{t-\Delta t} + \sqrt{\frac{\bm{\alpha}^2_{t-\Delta t}-\bm{\alpha}^2_t}{\bm{\alpha}^2_{t-\Delta t}}}\circ\bm{\xi}.
\end{equation}
Thus,
\begin{equation}
\bm{H}_t=\frac{\bm{\alpha}_t}{\bm{\alpha}_{t-\Delta t}}\circ \bm{H}_{t-\Delta t}+\sqrt{1-\left(\frac{\bm{\alpha}_t}{\bm{\alpha}_{t-\Delta t}}\right)^2}\circ\bm{\xi},
\end{equation}
which is a typical diffusion process.

Theorem \ref{the:forward} theoretically proves a simple idea: the final states are identical if the destination states are the same, regardless of the non-identical diffusion time steps that are selected. It becomes one theoretical insurance of non-identical diffusion models that we can directly investigate states instead of paths.
\subsection{Denoising process}
Similar to the denoising diffusion implicit models (DDIM) algorithm in identical diffusion, we have the following ODE and theorem, showing a typical non-identical DDIM process.
\begin{equation}\label{ODE_inverse}
    \d \bm{G}_t/\d t=[\log\bm{\beta}_{T-t}]'\circ\left(\bm{\alpha}^{-1}_{T-t}\circ\mathcal{D}^{\bm{\alpha}}_{T-t}(\bm{G}_t)-\bm{G}_t\right):=b_t^{\bm{\alpha}}(\bm{G}_t),
\end{equation}
where 
\begin{equation}
    \mathcal{D}^{\bm{\alpha}}_t(\bm{h}):=\mathbb{E}(\bm{H}_0|\bm{\alpha}_t\circ\bm{H}_0+\bm{\beta}_t\circ\bm{\xi}=\bm{h}).
\end{equation}
\begin{theorem}\label{the:DDIM} \textbf{(Correctness of non-identical DDIM)} Consider ODE \eqref{ODE_inverse} under the restrictions in Definition \ref{def:1}. Assume that $\mu$ is bounded and has compact support. With any $\bm{\alpha}_t$ satisfying the requirements in Definition \ref{def:1} and any starting point $t_0$, if $\bm{G}_{t_0}\sim\phi^{\bm{\alpha}}_{T-t_0}(\bm{g})$, we have
\begin{equation}\label{eq:DDIM}
    \bm{G}_t\sim\phi^{\bm{\alpha}}_{T-t}(\bm{g}), \forall t\in[t_0, T],
\end{equation}
and thus $\bm{G}_T\sim\mu$. Therefore, for any initialization and any given path under the requirements in Definition \ref{def:1}, the inverse generation result always follows the original distribution of the dataset.
\end{theorem}
\begin{proof}
    See Appendix \ref{app:proof}.
\end{proof}

Theorem \ref{the:DDIM} shows that if the starting points are the same, all denoising paths lead to the target distributions. Thus, the proposed formulation provides flexibility in choosing practical denoising paths for generation from the same starting distribution.

Correspondingly, substituting differentials with differences in \eqref{ODE_inverse}, we have the following vanilla non-identical DDIM algorithm.
\begin{equation}\label{DDIM_0}
\begin{aligned}
\bm{g}^\textrm{next}=&\left[\bm{\alpha}^\textrm{next}-\bm{\alpha}\circ\bm{\beta}^\textrm{next}\circ\bm{\beta}^{-1}\right]\circ\mathcal{D}^{\bm{\alpha}}(\bm{g})+\\&\bm{\beta}^\textrm{next}\circ\bm{\beta}^{-1}\circ\bm{g},
\end{aligned}
\end{equation}
which can be directly obtained by substituting the derivation in \eqref{ODE_inverse} by differentiation. Furthermore, as suggested in the DDIM paper \cite{ddim}, we actually use the following hybrid version of DDPM and DDIM for better generation.
\begin{equation}\label{DDIM}
\begin{aligned}
\bm{g}^\textrm{next}=&\left[\bm{\alpha}^\textrm{next}-\varepsilon\bm{\alpha}\circ\bm{\beta}^\textrm{next}\circ\bm{\beta}^{-1}\right]\circ\mathcal{D}^{\bm{\alpha}}(\bm{g})+\\
&\varepsilon\bm{\beta}^\textrm{next}\circ\bm{\beta}^{-1}\circ\bm{g}+\sqrt{1-\varepsilon^2}\bm{\beta}^\textrm{next}\bm{\xi},
\end{aligned}
\end{equation}
where the subscripts indicating time $t$ are omitted, $\bm{\xi}$ is an independently generated standard Gaussian noise and $\varepsilon$ is any constant between 0 and 1, which balances the generation quality and diversity.

Moreover, combining Theorems \ref{the:forward} and \ref{the:DDIM}, we can infer that it is not necessary to cover all the potential generation paths during the training phase. Instead, covering all the high-frequency $\bm{\alpha}_t$ values during training is necessary when designing the training algorithm, which presents a new challenge in non-identical diffusion systems.

The error occurs due to the approximation error between the NN and the expectation, as well as the error introduced by replacing differentials with differences. The second item relies on the original data distribution $\mu$, which is hard to analyze quantitatively. However, similar problems have been well studied in designing the generation steps of identical diffusion models, and these can be directly extended to the non-identical case. The unique challenge introduced by non-identity arises from the NN's approximation error. The error behavior of the NN depends on both the noise power and the pattern. By choosing different paths for $\bm{\alpha}_t$, we are effectively determining the evolution of the noise pattern. Even if the starting and ending points are fixed, we can still benefit from selecting an appropriate path. However, this problem is also infeasible to analyze quantitatively and is therefore discussed in the simulation section of this paper.

\section{Realizing Non-Identical Diffusion with NN}

In this section, we introduce the NN used for non-identical diffusion models, and we provide the training and generation algorithm. As is well known in identical diffusion models, the NN is used to fit $\mathcal{D}^{\bm{\alpha}}_t(\bm{h})$. We also note that the NN fitting can be indirect, such as using the velocity parameterization method \cite{v-pred1, v-pred2}, where a linear combination of the ground truth and the added noise becomes the raw output of the NN.

\subsection{Training}

For training stability, we normalize the input before feeding it into the NN. In identical diffusion, a common choice is $\widetilde{\bm{h}}:=\sqrt{d}\bm{h}/(\mathbb{E}\lVert\bm{h}\rVert_2^2)^{1/2}$, where $\bm{h}$ denotes a raw sample from the dataset. In wireless scenarios, we instead use $\widetilde{\bm{h}}:=\sqrt{d}\bm{h}/\lVert\bm{h}\rVert_2$ for better stability, since the sample power can usually be estimated from the received signal.

However, in non-identical diffusion methods, since we do not necessarily need to directly use the variable in diffusion as the NN's input, determining whether to highlight the more reliable elements becomes a new problem. Specifically, we consider the following two different methods of representing $\widetilde{\bm{h}}_t=\bm{\alpha}_t\circ\widetilde{\bm{h}}_0 + \bm{\beta}_t\circ\bm{\xi}$ for the NN input $\widehat{\bm{h}}_t$:
\begin{itemize}
    \item \textbf{Identical Total Power}: Here, we normalize the total power including original sample and the added noise of the input, that is, $\widehat{\bm{h}}_t:=\widetilde{\bm{h}}_t$.
    \item \textbf{Identical Noise Power}: To highlight the more reliable elements, we can also normalize the power of the added noise. That is, $\widehat{\bm{h}}_t:=Z\bm{\beta}_t^{-1}\circ\widetilde{\bm{h}}_t$, where $Z$ is the normalize factor to ensure $\|\widehat{\bm{h}}_t\|_2^2=d$. Recalling that $\bm{\alpha}^2+\bm{\beta}^2=\bm{1}$, by simple calculations we know $Z=\sqrt{d}\|\bm{\beta}_t^{-1}\|_2^{-1}$.
\end{itemize}
Generally, the identical total power input normalizes all elements in similar scales, which is usually beneficial for NN training. However, the noise components are amplified alongside the ground truth component, and thus the NN has to rely on the time embedding to obtain the non-identical error information. Meanwhile, the identical noise power scheme can correctly highlight the more accurate elements. However, the raw inputs may differ by several orders of magnitude, which usually harms NN representation.

\begin{algorithm}[t]
  \caption{Non-Identical Diffusion Model Training (Velocity Parameterization) \label{alg:train}}
  \begin{algorithmic}
  \STATE \textbf{Input:} Initialized NN $f(\bm{h}, \bm{\tau};\bm{\theta})$, batch size $B$, learning 
  \STATE rate $\eta$, maximum time step $T$, sampling method of $\bm{\tau}$
  \WHILE{not converged}
  \STATE Sample $\{\bm{h}_1,\cdots,\bm{h}_B\}$ from the dataset;
  \STATE Independently sample $\{\bm{\tau}_1,\cdots,\bm{\tau}_B\}$;
  \STATE Independently sample $\{\bm{\xi}_1,\cdot,\bm{\xi}_B\}$ from standard Gaussian distribution with the same shape as $\bm{h}$;
  \STATE $\bm{\alpha}_i\leftarrow\gamma(\bm{\tau}_i)$, where $\gamma$ is defined in \eqref{gamma};
  \STATE $\bm{\beta}_i\leftarrow\sqrt{1-\bm{\alpha}_i^2}$, $\widehat{\bm{h}}_i\leftarrow \bm{\alpha}_i\circ \bm{h}_i+\bm{\beta}_i\bm{\xi}_i$;
  \IF{using identical noise power}
  \STATE{$\widehat{\bm{h}}_i\leftarrow \sqrt{d}\|\bm{\beta}^{-1}_i\|^{-1}\bm{\beta}^{-1}_i\circ\widehat{\bm{h}}_i$};
  \ENDIF
  \STATE $\bm{y}_i\leftarrow\bm{\alpha}_i\circ\bm{\xi}_i-\bm{\beta}_i\circ\bm{h}_i$(Velocity Parameterization);
  \STATE $\ell\leftarrow \sum_{i=1}^B \|f(\widehat{\bm{h}}_i, \bm{\tau};\bm{\theta})-\bm{y}_i\|_2^2$;
  \STATE $\bm{\theta}\leftarrow\bm{\theta}-\eta\partial\ell/\partial\bm{\theta}$;
  \ENDWHILE
  \end{algorithmic}
\end{algorithm}
Before completing the training algorithm, we still need to determine the choice of $\bm{\alpha}_t$ during training, which is essential and can severely affect performance.
Unlike identical diffusion models, where we only have one diffusion path, we use a group of $\bm{\alpha}_t$ paths in non-identical diffusion models for generality.
Intuitively, the selection of representative $\bm{\alpha}_t$s should sufficiently cover the noise patterns used in generation to ensure successful results. Moreover, since we do not actually care about the entire diffusion process while adding noise, we only need to focus on the distribution of all $\bm{\alpha}_t$s when randomly selecting $\bm{\alpha}$ and $t$.

To simplify this procedure, we use a non-increasing unary function $\gamma: [0, T) \rightarrow (0,1]$, such that $\gamma(0)=1$, $\lim_{\tau\rightarrow T}\gamma(\tau)=0$, and $\gamma(\bm{\tau})$ follows the same distribution as $\bm{\alpha}_t$.
We note that the function $\gamma$ can be directly introduced from the common function $\alpha$ in identical diffusion models, where vector $\bm{\tau}$ plays the non-identical part. Therefore, we can use the following $\gamma$ function for integer inputs with the common hyperparameter setting \cite{ddpm} and define different noise patterns by selecting $\bm{\tau}$.
\begin{equation}
    \gamma(\tau):=\prod_{i=1}^\tau \sqrt{1 - 0.2i/T}.\label{gamma}
\end{equation}

Specifically, in this paper, we consider the following straightforward noise patterns for training defined through $\bm{\tau}$.
\begin{itemize}
    \item \textbf{Same}: A method to imitate identical diffusion, where $\bm{\tau}=x\bm{1}$ and $x$ are uniformly drawn from $\{0, \cdots, T-1\}$. The model downgrades to identical diffusion if only this method is applied.
    \item \textbf{Independent}: Another straightforward method where each element in $\bm{\tau}$ are independently and uniformly drawn from $\{0, \cdots, T-1\}$. It ensures extreme variations of the data distribution, while requiring more training effort to converge.
    \item \textbf{Pattern-Independent}: A method between same and completely independent drawing. It is designed to imitate the distribution of real-case raw estimations in the considered MIMO-OFDM system. Since the dimension $d$ is usually very large in practice, completely independent drawing becomes inefficient. By exploring the patterns in the potential initialization for generation, we can draw only a small set of $\bm{\tau}$ randomly and then complete it with a given pattern. In this paper, we use recurrent patterns since we usually insert pilots recurrently in OFDM systems.
    \item \textbf{Mixed}: For better generality, we can use a mixed method where we randomly choose noise patterns generated from the methods above.
\end{itemize}

Further, we denote the NN as $f(\bm{h}, \bm{\tau};\bm{\theta})$, where $\bm{\theta}$ is the parameter and take velocity parameterization \cite{v-pred1, v-pred2} as an example. Thus, the target of the NN becomes
\begin{equation}
\min_{\bm{\theta}} \textrm{KL}\left(f(\bm{H}, \bm{\tau};\bm{\theta})\|\gamma(\bm{\tau})\circ\bm{\xi}-\sqrt{1-\gamma^2(\bm{\tau})}\circ\widetilde{\bm{h}}_0\right),
\end{equation}
where KL$(\cdot\|\cdot)$ denotes the Kullback-Leibler Divergence. Furthermore, by variational inference, we can define the loss function as
\begin{equation}
    \ell= \sum_{i=1}^B \|f(\widehat{\bm{h}}_i, \bm{\tau};\bm{\theta})-\bm{y}_i\|_2^2,
\end{equation}
where $\bm{y}_i:=\gamma(\bm{\tau}_i)\circ\bm{\xi}_i-\sqrt{1-\gamma^2(\bm{\tau}_i)}\circ\widehat{\bm{h}}_i$.

\subsection{Generation}
\begin{algorithm}[t]
  \caption{Non-Identical Diffusion Generation \label{alg:gen}}
  \begin{algorithmic}
  \STATE \textbf{Input:} Trained NN $f(\bm{h}, \bm{\tau};\bm{\theta})$, initial estimation $\widetilde{\bm{h}}^{(0)}$, 
  \STATE  initial time indicator $\bm{\tau}^{(0)}$, total generation steps $N_\textrm{G}$
  \STATE $\bm{g}\leftarrow\widetilde{\bm{h}}^{(0)}$, $\bm{\tau}\leftarrow\bm{\tau}^{(0)}$;
  \STATE $\bm{\alpha}\leftarrow\gamma(\bm{\tau})$, $\bm{\beta}\leftarrow\sqrt{1-\bm{\alpha}\circ\bm{\alpha}}$, where $\gamma$ is defined in \eqref{gamma};
  \WHILE{$\bm{\tau}$ is not zero}
  \STATE Calculate $\bm{\tau}^\textrm{next}$ by \eqref{timegen};
  \STATE $\bm{\alpha}^\textrm{next}\leftarrow\gamma(\bm{\tau}^\textrm{next})$, $\bm{\beta}^\textrm{next}\leftarrow\sqrt{1-\bm{\alpha}^\textrm{next}\circ\bm{\alpha}^\textrm{next}}$;
  \IF{using identical noise power}
  \STATE $\bm{g}\leftarrow \sqrt{d}\|\bm{\beta}^{-1}\|^{-1}\bm{\beta}^{-1}\circ\bm{g}$;
  \ENDIF
  \STATE ${D}^{\bm{\alpha}}(\bm{g})\leftarrow\bm{\alpha}\circ\bm{g} - \bm{\beta}\circ f(\bm{g}, \bm{\tau};\bm{\theta})$ (Velocity Parameterization);
  \STATE Calculate $\bm{g}^\textrm{next}$ by \eqref{DDIM};
  \STATE $\bm{g}\leftarrow\bm{g}^\textrm{next}$, $\bm{\tau}\leftarrow\bm{\tau}^\textrm{next}$, $\bm{\alpha}\leftarrow\bm{\alpha}^\textrm{next}$;
  \ENDWHILE
  \STATE Output $\widetilde{\bm{h}}\leftarrow\bm{g}$ as final generation.
  \end{algorithmic}
\end{algorithm}
A typical generation procedure mainly follows \eqref{DDIM}. More specifically, we take the velocity parameterization as an example and show the generation procedure of one sample in Algorithm \ref{alg:gen}. However, as in the original DDIM algorithm, the time flow in generation does not necessarily need to be the same as in training. Especially when the initial time is non-identical, we generally have two generation strategies: uniform stepping and water-filling.

As described above, we use the same unary function $\gamma$ to map non-identical $\bm{\tau}$ to $\bm{\alpha}_t$. Again, since we usually care about the $\bm{\alpha}$ value after each step rather than the diffusion procedure itself (as in common identical DDIM systems), we focus on the evolution of $\bm{\tau}$ instead. Specifically, we denote $\bm{\tau}^{(0)}$ as the initial value obtained from the estimation algorithm, and $N_{\mr G}$ as the total number of generation steps.

The uniform stepping method is straightforward, where we reduce the time uniformly so that $\bm{\tau}^\textrm{next} = \bm{\tau} - \bm{\tau}^{(0)}/N_{\mr G}$.
Another straightforward method is to first improve the worse elements to the reliability of the better ones, and then proceed as in the identical DDIM, i.e., the water-filling algorithm. It always seeks to fill the gap between worse ones and better ones before improving them together. That is, we always try to find a $\bm{\tau}^\textrm{next}$ such that $\bm{\tau}^\textrm{next} \preceq \bm{\tau}$, $|\bm{\tau} - \bm{\tau}^\textrm{next}|_1 = |\bm{\tau}^{(0)}|_1/N_{\mr G}$, and $\bm{\tau}^\textrm{next}_i \leq \bm{\tau}^\textrm{next}_j$ always holds if $\bm{\tau}_i \leq \bm{\tau}_j$. In the following, we denote this procedure as $\bm{\tau}^\textrm{next} = \textrm{Waterfilling}(\bm{\tau}, |\bm{\tau}^{(0)}|_1/N_{\mr G})$.

We can also define a hybrid stepping method as:
\begin{equation}
\bm{\tau}^\textrm{next} = \epsilon(\bm{\tau} - \bm{\tau}^{(0)}/N_{\mr G}) + (1 - \epsilon)\textrm{Waterfilling}(\bm{\tau}, |\bm{\tau}^{(0)}|_1/N_{\mr G}),\label{timegen}
\end{equation}
where $\epsilon \in [0,1]$. This approach reduces to uniform stepping when $\epsilon = 1$, and to water-filling when $\epsilon = 0$, thereby completing Algorithm \ref{alg:gen}. Similarly, we can also define these stepping methods based on $\bm{\alpha}$ instead of $\bm{\tau}$. The details of the velocity parameterization method \cite{v-pred1, v-pred2} are omitted here due to page limitations.

\section{Non-Identical Time Embedding for Multi-Antenna OFDM Channels}
In this section, we tackle a technical problem: how to embed $\bm{\tau}$ into the NN. Unlike identical diffusion models, we have an independent time indicator for each element. However, common time embedding methods typically embed the scalar time into specific patterns, which cannot be directly applied to non-identical diffusion models. Thus, we propose averaging the current $\bm{\tau}$ along each dimension and performing embedding independently.

\begin{figure}
    \centering
    \includegraphics[width=0.95\linewidth]{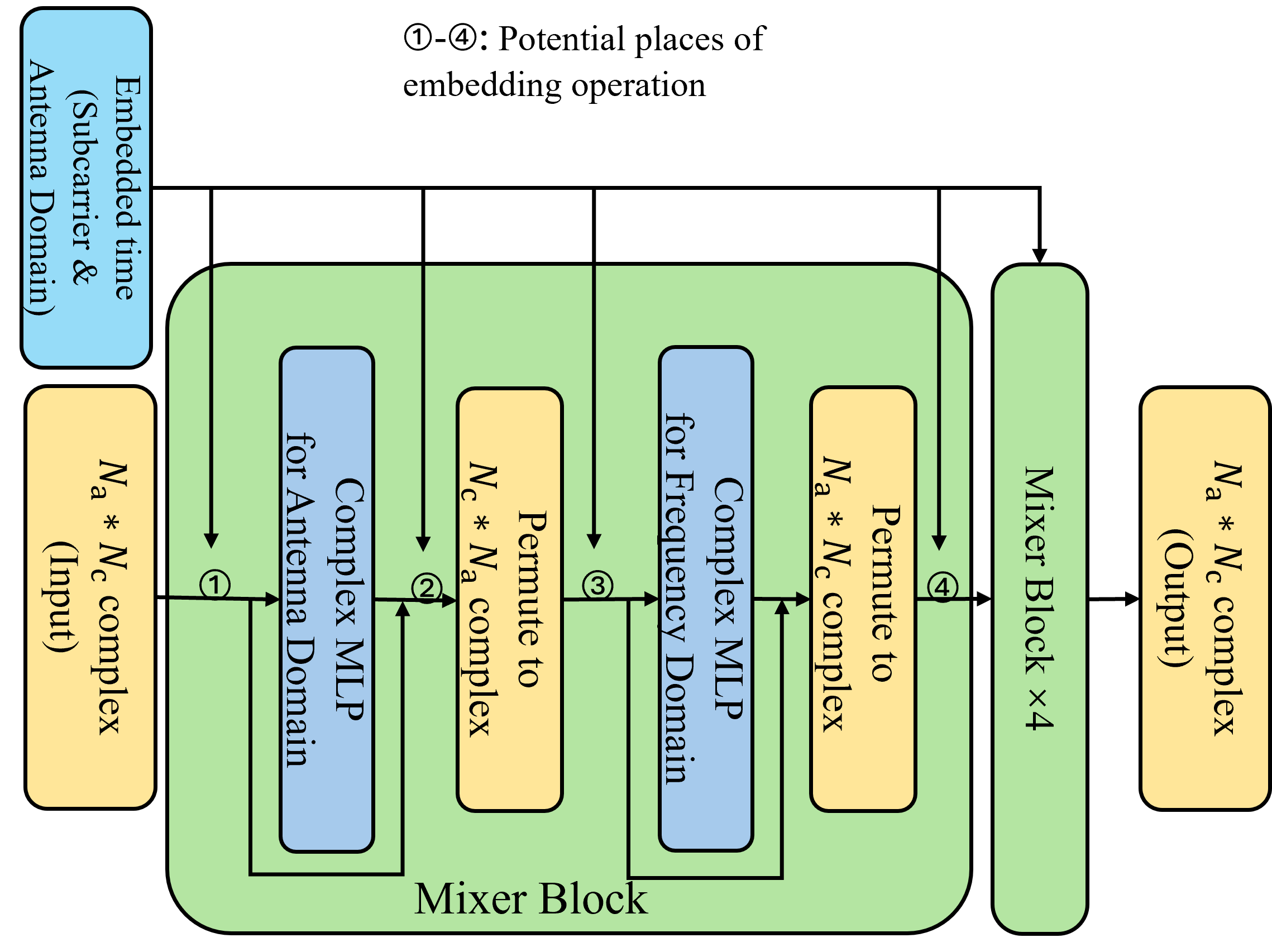}
    \caption{The backbone NN structure used in this paper, where \textcircled{1}-\textcircled{4} are potential places for time embedding operations.}
    \label{fig:NN}
\end{figure}
\subsection{Problem Restatement and Backbone NN}
Here, we call back to the specific problem introduced in Section II. Specifically, the problem can be stated as follows: Given $\widetilde{\mathbf{H}}$ and $\mathbf{A}$ in \eqref{init}, try to generate the ground truth $\mathbf{H}$. Such a problem can be applied to Algorithm \ref{alg:gen} by substituting $\tilde{\bm h}^{(0)}$ with $\widetilde{\mathbf{H}}$ and $\bm{\tau}^{(0)}$ with $\gamma^{-1}(\mathbf{A})$, respectively.
We note that although the channel here is a two-dimensional complex matrix instead of the real vectors discussed above, the difference only lies in the design of the backbone NN, and the algorithms described above can be directly applied.
For the backbone NN, we adopt the interleaved learning scheme based on MLP-Mixer, which has shown incredible performance in similar tasks \cite{mixer, 10942479, tnse}. Specifically, we adopt the exact same backbone NN as in these works with different time embedding methods, which is also illustrated in Fig. \ref{fig:NN}.

\subsection{Two-Dimensional Time Embedding}
\begin{figure}
    \centering
    \includegraphics[width=\linewidth]{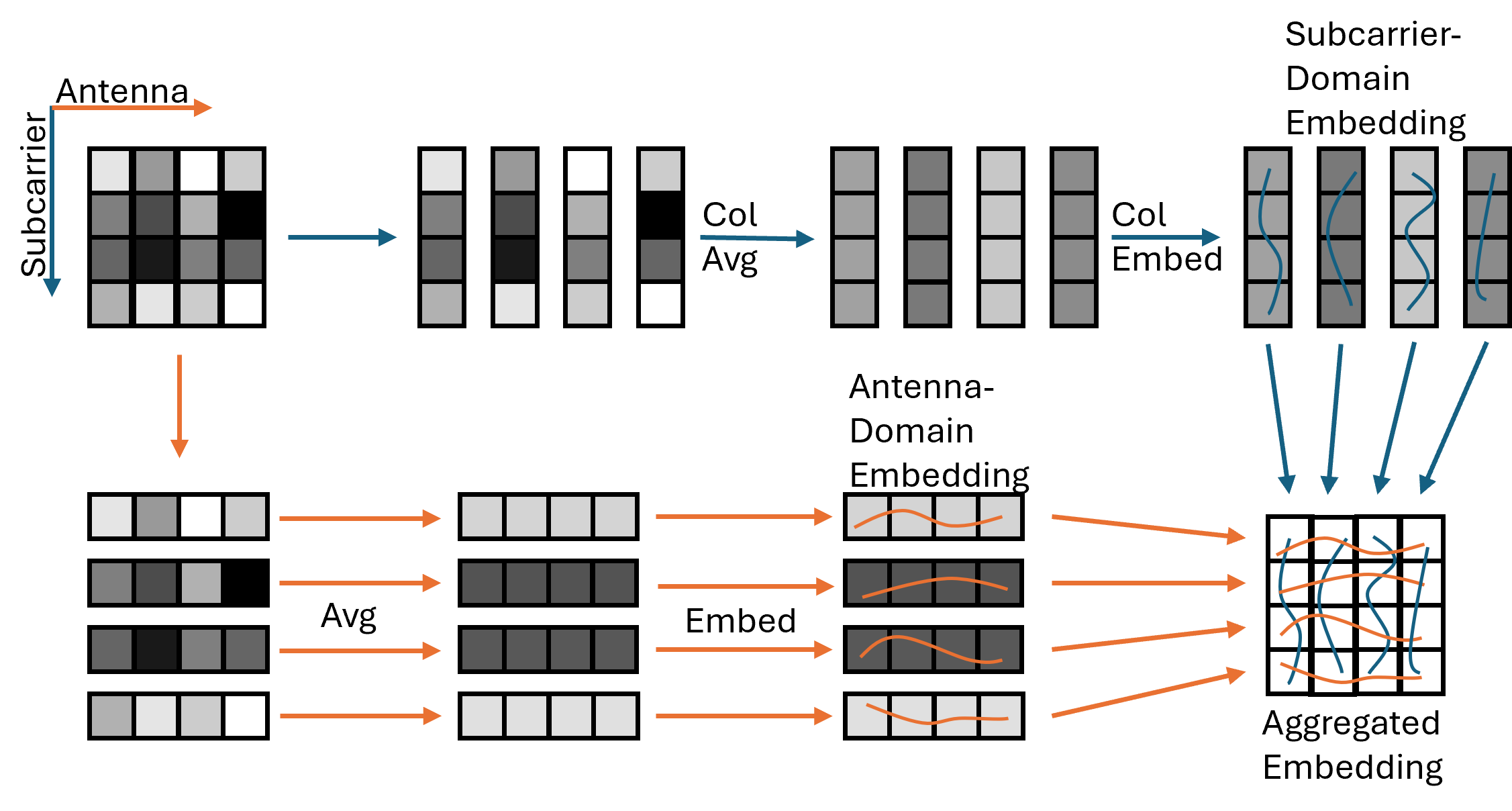}
    \caption{A conceptual illustration of the proposed two-dimensional time embedding method}
    \label{fig:embed}
\end{figure}
In non-identical diffusion models, conducting effective time embedding becomes a new problem. In identical diffusion models, we usually learn a mapping between unary time and patterns for embedding. However, in non-identical cases, each element corresponds to an independent time, making it impossible to generate a pattern for each element in the time matrix.

On the other hand, we note that in multi-antenna OFDM systems, the resource allocation is typically based on subcarriers or antennas, inspiring us to conduct time embedding per subcarrier and per antenna. By averaging along each dimension, the averaged result can maintain most of the original characteristics across almost all pilot schemes. Moreover, such a one-dimensional embedding structure can be easily incorporated into the interleaved learning scheme, as illustrated in Fig. \ref{fig:embed}.

In particular, in the interleaved learning scheme, we have two groups of modules dedicated to subcarriers and antennas, respectively. From the time embedding scheme, we can also generate two time embedding vectors corresponding to both dimensions. Intuitively, we can use the following averaged time for embedding.
\begin{eqnarray}
\bm{\tau}^\textrm{(time)}_{\mr c} &= \textrm{round}(\bm{\tau}\bm{1}/N_{\mr c}),\\
\bm{\tau}^\textrm{(time)}_{\mr a} &= \textrm{round}(\bm{\tau}^T\bm{1}/N_{\mr a}),
\end{eqnarray}
where round$(\cdot)$ denotes the rounding function producing integer time for further processing. We note that $\bm{\tau}_{\mr c}$ averages the time in different subcarriers and keeps each antenna independent. Therefore, it is called subcarrier-domain mapping. Similarly, $\bm{\tau}_{\mr a}$ is called the antenna-domain mapping.

On the other hand, as the time in a diffusion model stands for the power ratio of the desired variable in the estimation, we can also average over $\bm{\alpha}$ for power-averaging.
\begin{eqnarray}
\bm{\tau}^\textrm{(power)}_{\mr c} =& \textrm{round}(\gamma^{-1}\left(\gamma(\bm{\tau})\bm{1}/N_{\mr c})\right),\\
\bm{\tau}^\textrm{(power)}_{\mr a} =& \textrm{round}(\gamma^{-1}\left(\gamma(\bm{\tau})^T\bm{1}/N_{\mr a})\right),
\end{eqnarray}

Incorporating this into the interleaved learning framework, we have the following three embedding methods, which complete the NN design.
\begin{itemize}
    \item \textbf{Row-Wise Embedding:} We perform time embedding along the same domain as the interleaved learning unit. That is, we apply the subcarrier-domain embedding $\bm{\tau}_{\mr a}$ at points \textcircled{1} and \textcircled{2}, and $\bm{\tau}_{\mr c}$ at \textcircled{3} and \textcircled{4} in Fig. \ref{fig:NN}, respectively.
    \item \textbf{Column-Wise Embedding:} We can also perform time embedding along the opposite domain of the interleaved learning unit. In other words, the antenna-domain embedding $\bm{\tau}_{\mr a}$ is applied to the subcarrier-processing blocks, while the subcarrier-domain embedding $\bm{\tau}_{\mr c}$ is applied to the antenna-processing blocks.
    \item \textbf{Embedding Together:} We can also overlook the detailed structure of the interleaved learning and conduct both embeddings integrally. That is, we apply both $\bm{\tau}_{\mr c}$ and $\bm{\tau}_{\mr a}$ at points \textcircled{1} and \textcircled{4}, and do nothing at \textcircled{2} and \textcircled{3} in Fig. \ref{fig:NN}.
\end{itemize}
\subsection{Complexity of the Two-Dimensional Time Embedding}
Here, we discuss the additional complexity introduced by the proposed two-dimensional time embedding structure based on the network structure shown in Fig. \ref{fig:NN}. We note that in traditional identical diffusion models, we only conduct row-wise embedding as the column-wise embedding downgrades to adding a constant.

As the total time steps are the same in identical and non-identical diffusion systems, the embedding matrices are of the same size.
Therefore, the number of parameters remains constant when introducing the non-identical diffusion structure, while this adds more computational complexity. 

According to the embedded position, we have two sizes of the diffusion head corresponding to $N_\mr{a}$ and $N_\mr{c}$, whose computational burdens are proportional, denoted by $N_\mr{a}C$ and $N_\mr{c}C$, respectively.
For all variations of diffusion models mentioned in this section, we apply each diffusion head once every layer. In the identical case, we only have a scalar time identifier. Thus, the embedding computational complexity of each layer becomes $(N_\mr{a}+N_\mr{c})C$. On the other hand, in the non-identical cases, we have an additional batch size of $N_\mr{c}$ for antenna-domain embedding, and $N_\mr{a}$ for the subcarrier-domain one, respectively. Thus, the embedding computational complexity of each layer becomes $2N_\mr{a}N_\mr{c}C$.

Also, we note that the proposed non-identical diffusion scheme also introduces more computational complexity in averaging time steps, time scheduling, and generation steps. However, such operations are non-parametric, and the computational overhead can be overlooked compared with that of the NN.

Generally, introducing the non-identical diffusion model scheme does not introduce any extra trainable parameters. The computational complexity of the NN propagation part also remains the same. However, the computational burden regarding the time embedding becomes $2N_\mr{a}N_\mr{c}/(N_\mr{a}+N_\mr{c})$ times. There are also some other extra computational overheads which are not comparable to existing ones.
\section{Numerical Results}
\subsection{Dataset and Basic Settings}
Following the experimental settings in \cite{tnse}, we consider the multi-antenna OFDM scenario using the DeepMIMO \cite{deepmimo} dataset. We use the same data preprocessing method as in \cite{tnse}. The shape of the channel matrix $\mathbf{H}$ is given by $N_{\mr a}=32$, $N_{\mr c}=64$, corresponding to the number of BS antennas and subcarriers. The first subcarrier lies at 3.5 GHz, and the subcarrier spacing is 300 kHz. We have 43,440 channel matrices in the training dataset and 10,860 in the testing dataset. In the training phase, we only use the training dataset, while the testing dataset is exclusively used in the generation experiments. We note that we omit the detailed communication scheme in this paper and only focus on how to generate high-quality channel matrices from partial, noisy initial observations.

Specifically, we show the computational cost (in terms of MACs) and parameter count for different parts of the utilized identical and non-identical models as follows. Each block of the backbone complex mixer network costs 1.61 GMACs and has 79.0K parameters. The embedding is conducted in two stages. The first two layers, with 33.2K parameters, are reused by all layers. They take 75.5 MMACs with identical diffusion and 1.61 GMACs with non-identical diffusion. The layer-specific embedding head with 37.4K parameters takes 37.8 MMACs and 0.805 GMACs for identical and non-identical schemes, respectively. 
Considering all 5 layers in general, we know that
\begin{itemize}
    \item In both identical and non-identical diffusion schemes, the total amount of parameters is 615K, of which 35.8\% are related to time embedding.
    \item In the identical diffusion scheme, one propagation over a channel sample takes 8.31 GMACs, of which 3.2\% is related to time embedding.
    \item In the non-identical diffusion scheme, one propagation over a channel sample takes 13.87 GMACs, of which 41.2\% are related to time embedding. 
\end{itemize}
Generally, the proposed non-identical scheme brings about 67\% more computation burden caused by the time-embedding layers. However, due to the parallelization in GPU computing, it does not always cost that much of GPU computing time when the batch size is small as indicated in \cite{tnse}.
\subsection{Training Results}
\begin{table*}[t]\scriptsize
\begin{tabular}{|cc|cc|cc|cc|cc|}
\hline
\multicolumn{2}{|c|}{\multirow{2}{*}{}}                 & \multicolumn{2}{c|}{Same}                              & \multicolumn{2}{c|}{Independent}                       & \multicolumn{2}{c|}{Periodical}                        & \multicolumn{2}{c|}{Car-only}                          \\ \cline{3-10} 
\multicolumn{2}{|c|}{}                                  & \multicolumn{1}{c|}{Identical noise} & Identical power & \multicolumn{1}{c|}{Identical noise} & Identical power & \multicolumn{1}{c|}{Identical noise} & Identical power & \multicolumn{1}{c|}{Identical noise} & Identical power \\ \hline
\multicolumn{1}{|c|}{\multirow{2}{*}{Row}}  & $\bm{\tau}$ avg  & \multicolumn{1}{c|}{0.346 ± 0.024}   & 0.334 ± 0.022   & \multicolumn{1}{c|}{0.274 ± 0.001}   & 0.275 ± 0.001   & \multicolumn{1}{c|}{0.233 ± 0.009}   & 0.264 ± 0.006   & \multicolumn{1}{c|}{0.094 ± 0.002}   & 0.089 ± 0.002   \\ \cline{2-10} 
\multicolumn{1}{|c|}{}                      & $\bm{\alpha}$ avg & \multicolumn{1}{c|}{0.345 ± 0.024}   & 0.334 ± 0.022   & \multicolumn{1}{c|}{0.275 ± 0.001}   & 0.277 ± 0.001   & \multicolumn{1}{c|}{0.230 ± 0.009}   & 0.263 ± 0.006   & \multicolumn{1}{c|}{0.097 ± 0.003}   & 0.084 ± 0.002   \\ \hline
\multicolumn{1}{|c|}{\multirow{2}{*}{Col}}  & $\bm{\tau}$ avg  & \multicolumn{1}{c|}{0.331 ± 0.023}   & 0.327 ± 0.022   & \multicolumn{1}{c|}{0.265 ± 0.001}   & 0.277 ± 0.001   & \multicolumn{1}{c|}{0.228 ± 0.009}   & 0.262 ± 0.006   & \multicolumn{1}{c|}{0.072 ± 0.003}   & 0.068 ± 0.002   \\ \cline{2-10} 
\multicolumn{1}{|c|}{}                      & $\bm{\alpha}$ avg & \multicolumn{1}{c|}{0.334 ± 0.023}   & 0.327 ± 0.022   & \multicolumn{1}{c|}{0.262 ± 0.001}   & 0.280 ± 0.001   & \multicolumn{1}{c|}{0.222 ± 0.008}   & 0.258 ± 0.007   & \multicolumn{1}{c|}{0.070 ± 0.003}   & 0.068 ± 0.002   \\ \hline
\multicolumn{1}{|c|}{\multirow{2}{*}{Both}} & $\bm{\tau}$ avg  & \multicolumn{1}{c|}{0.338 ± 0.024}   & 0.318 ± 0.023   & \multicolumn{1}{c|}{0.326 ± 0.001}   & 0.259 ± 0.001   & \multicolumn{1}{c|}{0.232 ± 0.007}   & 0.248 ± 0.006   & \multicolumn{1}{c|}{0.062 ± 0.002}   & 0.057 ± 0.001   \\ \cline{2-10} 
\multicolumn{1}{|c|}{}                      & $\bm{\alpha}$ avg & \multicolumn{1}{c|}{0.351 ± 0.032}   & 0.318 ± 0.023   & \multicolumn{1}{c|}{0.316 ± 0.001}   & 0.263 ± 0.001   & \multicolumn{1}{c|}{0.223 ± 0.007}   & 0.244 ± 0.006   & \multicolumn{1}{c|}{0.059 ± 0.002}   & 0.055 ± 0.001   \\ \hline
\end{tabular}
\caption{Mean $\pm$ standard deviation of the converged training NMSE (epochs 51--100) under different non-identical training configurations.}\label{table:train}
\end{table*}
\begin{figure}[t]
    \centering
    \includegraphics[width=0.95\linewidth]{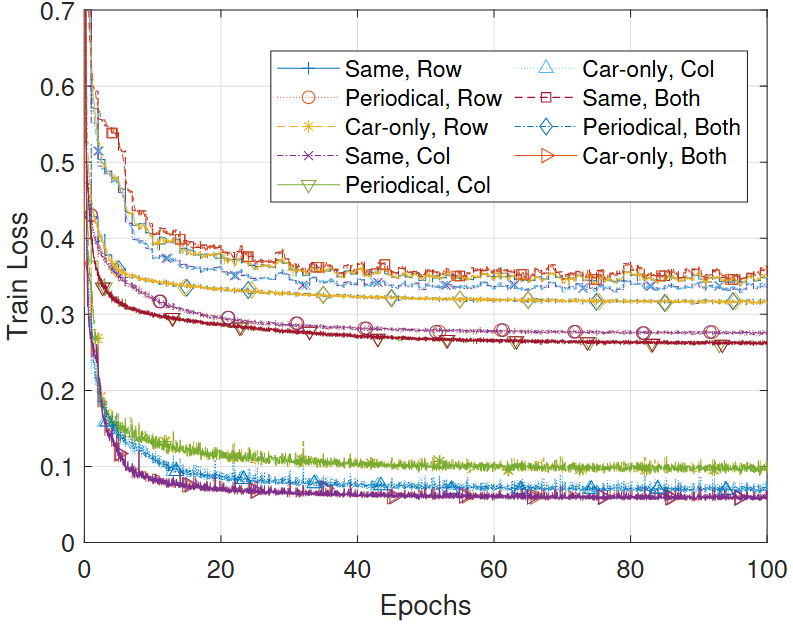}
    \caption{Training NMSE versus epoch for different time-embedding methods under $\alpha$-average sampling and identical-noise-power normalization.}
    \label{fig:train1}
\end{figure}
\begin{figure}[h]
    \centering
    \includegraphics[width=0.95\linewidth]{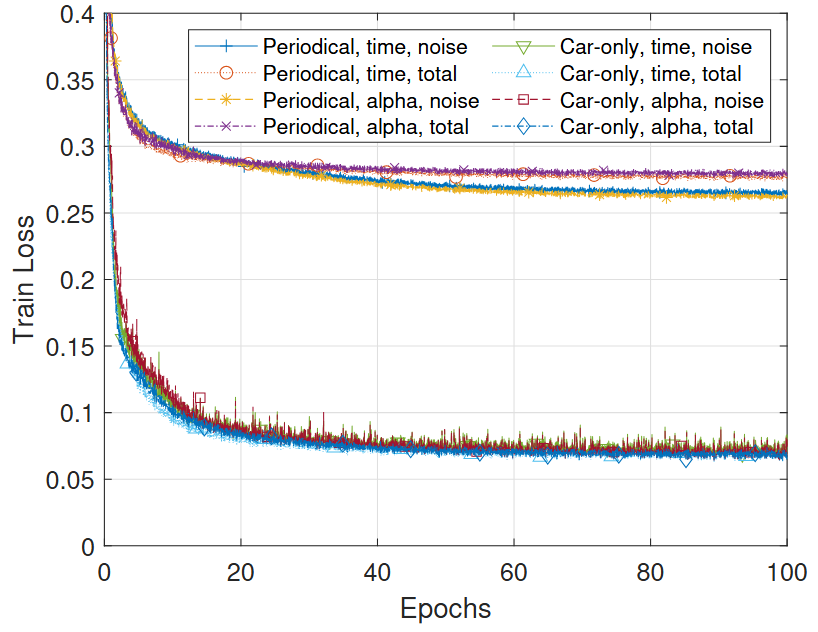}
    \caption{Training NMSE versus epoch for different normalization and averaging methods using column-wise time embedding.}
    \label{fig:train2}
\end{figure}
In this section, we show the training results of the proposed non-identical diffusion model. Specifically, we conducted 48 independent training runs, corresponding to two input normalization methods, three time-embedding schemes, two dimensional averaging methods, and four training noise patterns:
\begin{itemize}
    \item \textbf{Same:} the setting introduced in Section~IV-A, where all elements share the same sampled timestep;
    \item \textbf{Independent:} the setting introduced in Section~IV-A, where all elements independently sample their timesteps;
    \item \textbf{Periodical:} a structured pattern in which the noise matrix is periodic along both dimensions to imitate common pilot schemes; the period is uniformly sampled from 4 to 20 in the subcarrier domain and from 4 to 10 in the antenna domain;
    \item \textbf{Car-Only:} a structured pattern in which all antennas of the same subcarrier share the same noise power, while different subcarriers are independent, which is typical in uplink scenarios.
\end{itemize}

In the remainder of this section, we show some typical curves to compare different settings, and a complete result table of all 48 experiments can be found in Table \ref{table:train}. In addition to the experiments above, we also conducted experiments on two mixed noise patterns: the ``non-directional'' pattern, where we uniformly and randomly choose from patterns 1–3 above; and the ``all'' pattern, where we uniformly and randomly choose from all simple patterns.
In Table \ref{table:train}, we focus on the training loss from epochs 51 to 100 after convergence and show the mean value and standard deviation for each independent experiment. We note that it is not fair to compare the results across different noise patterns. Instead, we only focus on comparisons within the same noise patterns here, and we evaluate different training noise patterns through the generation results.

We also show the convergence curves for some specific experiments in Figs. \ref{fig:train1} and \ref{fig:train2}. In Fig. \ref{fig:train1}, we show the results under different time embedding methods with alpha-average sampling, and the input is normalized by identical noise power.
In Fig. \ref{fig:train2}, we show the results with different normalization and dimensional averaging methods using column-wise time embedding.
From Figs.~\ref{fig:train1} and \ref{fig:train2} and Table~\ref{table:train}, we make three observations. First, the Same pattern yields the highest training loss, whereas the Car-Only pattern is the easiest to fit because of its strong structure. Second, column-wise embedding generally achieves better converged performance than row-wise embedding. Third, normalizing by noise power improves the final converged loss but slows down early training because it makes the input scale less balanced. The two averaging strategies lead to only marginal differences in training loss.

\subsection{Generation Results}
\subsubsection{Generation Experiment Settings}
\begin{figure}[h]
    \centering
    \includegraphics[width=\linewidth]{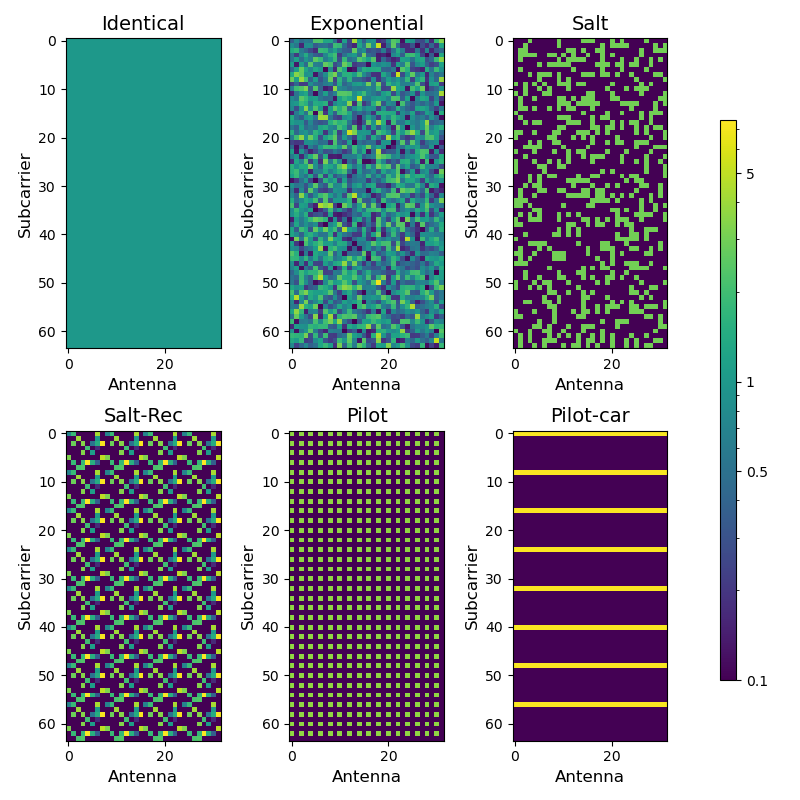}
    \caption{An illustration of different initial noise patterns. The noise power is normalized, and only the power map is shown.}
    \label{fig:noisepattern}
\end{figure}
When evaluating the generation performance, we use the test set of the aforementioned dataset. Specifically, we apply $N_{\mr G}=50$ steps of non-identical DDIM generation with $\varepsilon=0.4$. Within each experiment, we use the same noise pattern for the initial channel estimation. Specifically, we use the following noise initializations, which are also illustrated in Fig. \ref{fig:noisepattern}. We first allocate 10\% of the total noise energy evenly among all elements as background noise. The remaining 90\% is generated from one of the following distributions, completing the power map. The noise is independently Gaussian-distributed with the generated power map. We note that when using masked inputs, the following noises are only applied to the remaining elements.

\begin{itemize}
\item \textbf{White:} As a benchmark method, the power is identical among all elements, i.e., Gaussian white noise. The default signal-to-noise ratio (SNR) is -10 dB.
\item \textbf{Exponential (Exp):} The noise power of each element independently follows an exponential distribution with a rate of 0.5. The default SNR is -5 dB.
\item \textbf{Salt:} We apply a randomly generated mask retaining 30\% of the elements. The power is identical among all remaining elements. The default SNR is 0 dB.
\item \textbf{Salt-Rec:} The noise map is generated recurrently every 8 antennas and 8 subcarriers. Within each section, a mask retains 30\% of the elements, and the noise power is exponentially distributed with a rate of 0.5. Such a noise map is representative of actual pilot schemes, which are typically recurrent. The default SNR is 0 dB.
\item \textbf{Pilot:} We use a typical recurrent pilot scheme where the spacing is 2 in both antennas and subcarriers. The power is identical among all remaining elements. The default SNR is 10 dB.
\item \textbf{Pilot-Car:} In some applications like uplink transmission, all antennas are used simultaneously. Thus, we use another pilot setting where the period is 8 in subcarriers, and the antennas are either all masked or all retained for each subcarrier. The power is identical among all remaining elements. The default SNR is 10 dB.
\end{itemize}

\subsubsection{Comparison With Identical Diffusion}
\begin{figure}
    \centering
    \includegraphics[width=0.95\linewidth]{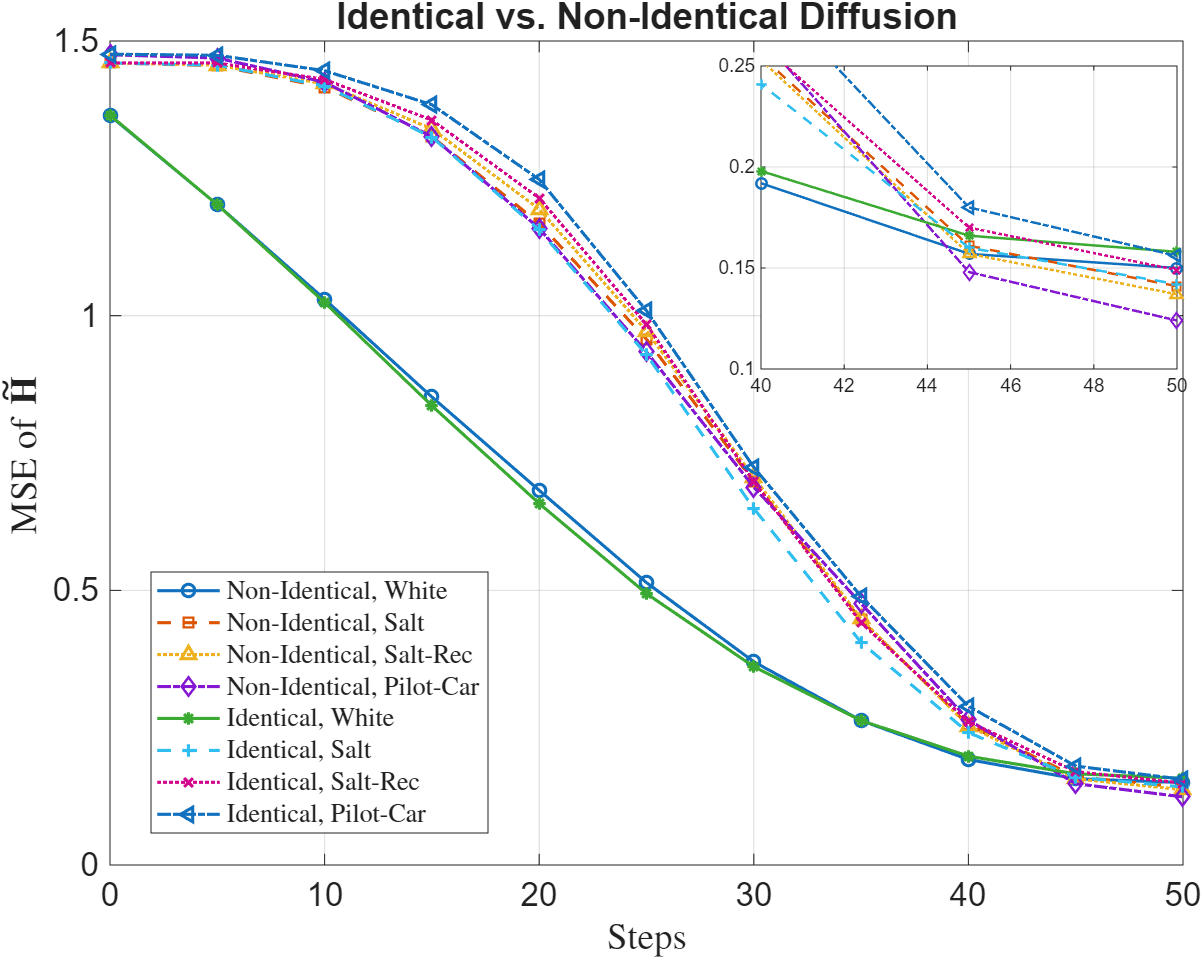}
    \caption{NMSE versus generation step for identical and non-identical diffusion under different initial noise patterns.}
    \label{fig:comp}
\end{figure}
In Fig. \ref{fig:comp}, we show the comparison between the proposed non-identical diffusion model and the traditional identical framework as in \cite{tnse}. We use four kinds of noise initialization for generation and show the NMSE error of the diffusion variable, i.e., channel matrix $\mathbf H$, after each step. In the non-identical diffusion settings, we use column-wise embedding, averaging over $\bm{\alpha}$, diffusion variables with normalized power, and a $\bm{\tau}$-waterfilling scheduler, which is chosen based on prior experiments.

Figure~\ref{fig:comp} shows that both the identical and non-identical variants progressively reduce the channel NMSE from the coarse initialization. Their performance is close for White and Salt initializations, while the proposed non-identical scheme performs better for structured non-uniform initializations, especially Pilot-Car and, to a lesser extent, Salt-Rec. This behavior is consistent with the motivation of the proposed time-map design: its advantage becomes more visible when the initial reliability is strongly heterogeneous across the channel matrix.
\subsubsection{Impact of Stepping Method}
\begin{table}[t]
    \centering\scriptsize
    \begin{tabular}{|c|c|c|c|c|c|}
        \hline
        Initialization & \textbf{Exp} & \textbf{Salt} & \textbf{Salt-Rec} & \textbf{Pilot} & \textbf{Pilot-Car} \\ \hline
        $\bm{\tau}$-Linear & 0.098 & 0.202 & 0.172 & 0.386 & 0.276 \\ \hline
        $\bm{\tau}$-Hybrid (0.7) & 0.095 & 0.189 & 0.155 & 0.367 & 0.225 \\ \hline
        $\bm{\tau}$-Hybrid (0.5) & 0.094 & 0.182 & 0.147 & 0.358 & 0.205 \\ \hline
        $\bm{\tau}$-Hybrid (0.3) & 0.093 & 0.177 & 0.141 & 0.350 & 0.193 \\ \hline
        $\bm{\tau}$-Waterfilling & 0.092 & \textbf{0.170} & \textbf{0.135} & \textbf{0.342} & \textbf{0.187} \\ \hline
        $\bm{\alpha}$-Linear & 0.096 & 0.230 & 0.199 & 0.389 & 0.295 \\ \hline
        $\bm{\alpha}$-Hybrid (0.7) & 0.093 & 0.213 & 0.177 & 0.373 & 0.231 \\ \hline
        $\bm{\alpha}$-Hybrid (0.5) & 0.092 & 0.205 & 0.170 & 0.367 & 0.218 \\ \hline
        $\bm{\alpha}$-Hybrid (0.3) & 0.091 & 0.199 & 0.164 & 0.363 & 0.214 \\ \hline
        $\bm{\alpha}$-Waterfilling & \textbf{0.090} & 0.192 & 0.158 & 0.361 & 0.219 \\ \hline
    \end{tabular}
    \caption{The last-step performance of different stepping methods under different initializations, where the numbers in blankets represent the value of $\epsilon$ in \eqref{timegen}.}
    \label{table:2}
\end{table}
Table \ref{table:2} shows the final generation results with different stepping methods. We use the column-wise embedding method with identical noise input and embedding based on $\bm{\alpha}$, which performs best during training. For the sake of generalization, we use the NN trained under the “All” noise pattern.
Comparing within each row, we observe that the waterfilling method based on time $\bm{\tau}$ outperforms all other stepping methods in most cases. This result is intuitive, since the NN usually works better when the noise distribution is more balanced, and existing works in identical diffusion models also show that stepping by $\bm{\tau}$ is superior to stepping by $\bm{\alpha}$. We also used the trained model with identical power inputs for the same experiments, which also showed similar results and is omitted here due to page limits.
\subsubsection{Generating Performance Under Different Training Noise Patterns}
\begin{table}[t]
    \centering\scriptsize
    \begin{tabular}{|c|c|c|c|c|c|c|}
        \hline
        Training Noise & \textbf{White} & \textbf{Exp} & \textbf{Salt} & \textbf{Salt-Rec} & \textbf{Pilot} & \textbf{Pilot-Car} \\ \hline
        Same & 0.263 & 0.100 & 0.328 & 0.306 & 0.817& 0.336\\ \hline
        Independent & 0.434 & 0.189 & 0.631 & 0.595 & 0.893& 0.628\\ \hline
        Periodical & 0.299 & 0.177 & 0.339 & 0.284 & 0.495&0.297 \\ \hline
        Car-Only & 1.028 & 0.715 & 1.076 & 0.950 & 1.069& 0.928\\ \hline
        Non-Directional & 0.171 & 0.100 & 0.182 & 0.146 & 0.453& 0.196\\ \hline
        All & \textbf{0.156} & \textbf{0.092} & \textbf{0.170} &\textbf{0.135} & \textbf{0.342}&\textbf{0.187} \\ \hline
    \end{tabular}
    \caption{The performance of different training noise patterns under different initialization of generation.}
    \label{table:3}
\end{table}
In this section, we investigate the effect of different training noise schemes, as shown in Table \ref{table:3}. We inherit the basic settings above and use the $\bm{\tau}$-Waterfilling method for stepping. As the noise pattern has a vital effect on the NN’s performance, aligning the noise-adding method during training with the initialization of generation becomes necessary. From Table \ref{table:3}, we observe the following interesting findings. Like what we did above, we also repeated the experiments with the trained model having identical power inputs, showing similar results, and also omit them here.

First, the “same” noise pattern, indicating identical training noise power, performs surprisingly well compared with most specific training patterns, verifying the success of previous identical diffusion works. Especially when the initial noise does not contain much structural information, it performs almost the best. This is intuitive since we cannot benefit much from the initial noise structure.
Meanwhile, although the models trained under specific noise patterns seem to work well during training, they fail to obtain good generation results—even under a similar initial noise pattern—except for the periodic training noise pattern. A possible explanation for this phenomenon is as follows. It is hard to describe such distributions by sampling due to their complexity, while all samplings share the same pattern. Thus, the NN may focus on some biased knowledge instead of the dimension-wise noise embedding, resulting in degraded generalization performance. Although such special noise patterns cannot lead to a good training result independently, they can still greatly improve the generation capability by providing diverse and complementary training signals, as inferred from the results of hybrid training noise patterns.

\subsubsection{Impact of Embedding Schemes}
\begin{figure}[h]
    \centering
    \includegraphics[width=0.95\linewidth]{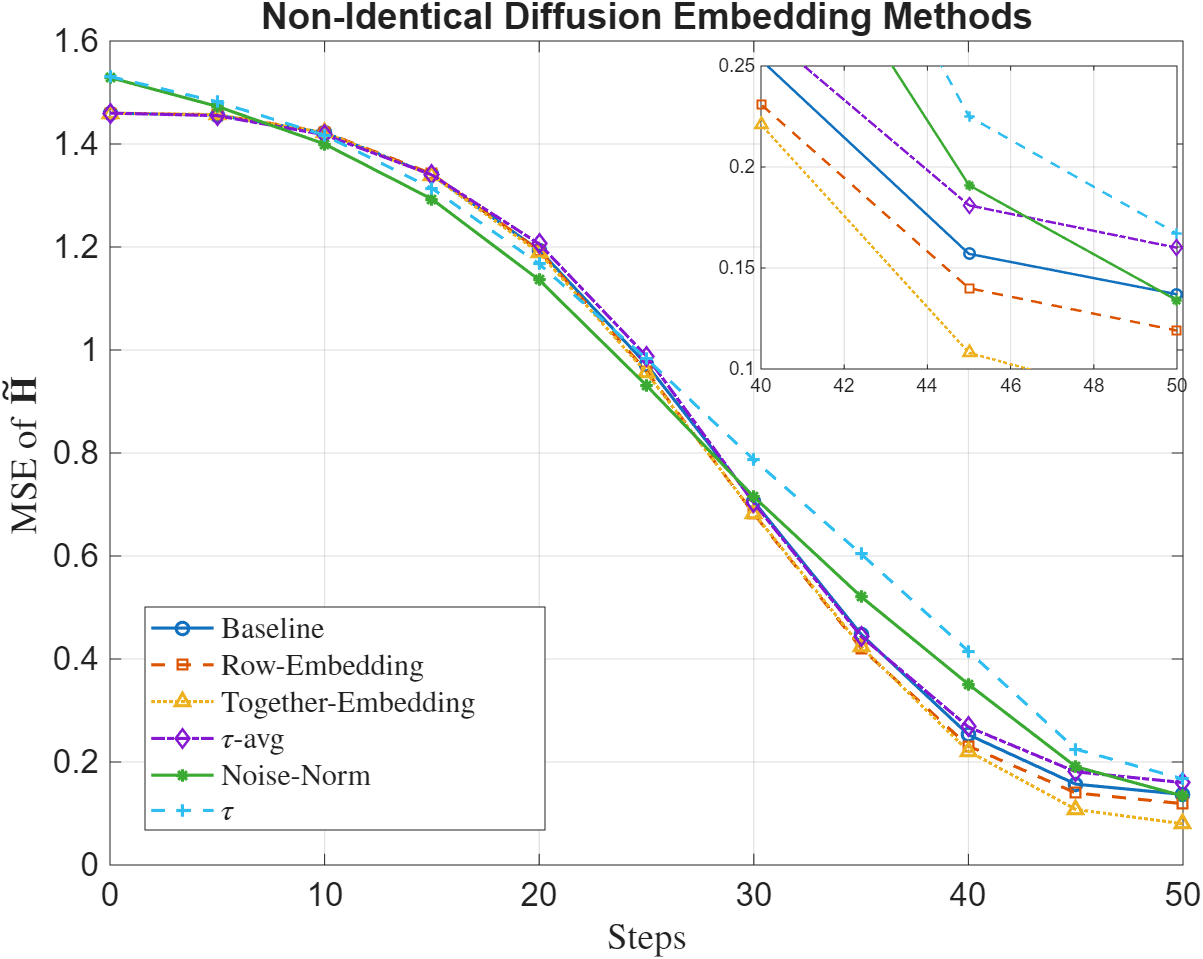}
    \caption{Final-generation NMSE under different embedding schemes. ``Baseline'' denotes column-wise embedding with $\alpha$-averaging and identical-noise-power normalization.}
    \label{fig:gentrain}
\end{figure}
Figure~\ref{fig:gentrain} reports the effect of different embedding schemes. Here, ``baseline'' denotes the setting with column-wise embedding, $\bm{\alpha}$-averaging, and identical-noise-power normalization, while the other labels indicate the corresponding deviations from this default configuration. However, the differences among embedding variants are relatively modest and can vary with the training and generation settings. Therefore, we use these results mainly to select a default configuration rather than to claim a universally superior embedding design.



\section{Conclusion and Future Directions}
In this paper, we introduced the non-identical diffusion model, which uses a matrix-based time representation to characterize element-wise heterogeneous noise levels. We presented the corresponding denoising formulation together with practical training and generation strategies. For the MIMO-OFDM channel generation task, we further proposed a dimension-wise time embedding scheme. Numerical results showed that the proposed framework is effective, especially when the initial channel estimate exhibits strongly non-uniform reliability across elements.

This paper mainly focused on the systematic design and proposed heuristic algorithms, leaving various directions for future work. One important direction is to further delve into the stepping methods. Our results showed that the waterfilling method usually outperforms other methods. Further efforts are needed to clarify whether this phenomenon also holds in other problems and if waterfilling indicates the optimal path in general non-identical diffusion models. Moreover, there seems to be a complicated interplay between the embedding schemes and training and generation performances, requiring further examination. In addition, the training method warrants further improvement, as our results indicate that the diversity of training data plays a critical role in generation performance.
For the MIMO-OFDM scenario examined, it is still essential to further combine the proposed methods with the actual resource block structures and signal modulations as in \cite{10942479, tnse}.

The code and additional results are available at \url{https://github.com/yuzhiyang123/Non-Identical_diffusion}.
\begin{appendices}
\section{Detailed Proof to Theorem \ref{the:forward}}\label{app:proof1}
Here we show the detailed derivation from \eqref{SDE_forward} to \eqref{SDE_eq}.
We define
\(\bm{X}_t := \bm{\alpha}_t^{-1}\circ \bm{H}_t.\)
Applying It\^o's product rule componentwise, and noting that there is no quadratic
variation term involving $\alpha_t^{-1}$ because it is deterministic and of finite
variation, we obtain
\begin{align}
\d \bm{X}_t
=&
(\bm{\alpha}_t^{-1})' \circ \bm{H}_t\d t + \bm{\alpha}_t^{-1}\circ \d\bm{H}_t \notag\\
=&
-(\log \bm{\alpha}_t)'\circ\bm{\alpha}_t^{-1}\circ \bm{H}_t\d t\notag\\
&+
\bm{\alpha}_t^{-1}\circ
\left(
(\log \bm{\alpha}_t)'\circ \bm{H}_t\d t
+
\sqrt{-2(\log \bm{\alpha}_t)'}\circ \d\bm{W}_t
\right) \notag\\
=&
\alpha_t^{-1}\circ \sqrt{-2(\log \bm{\alpha}_t)'}\circ dW_t.
\label{eq:Yt_sde}
\end{align}
Integrating \eqref{eq:Yt_sde} from $0$ to $t$ and using $\bm{\alpha}_0=\bm{1}$ gives
\begin{equation}
\bm{X}_t
=
\bm{H}_0
+
\int_0^t
\bm{\alpha}_s^{-1}\circ \sqrt{-2(\log \bm{\alpha}_s)'}\circ \d\bm{W}_s.
\end{equation}
Therefore,
\begin{equation}
\bm{H}_t
=
\bm{\alpha}_t\circ \bm{H}_0
+
\bm{Z}_t,
\bm{Z}_t
:=
\bm{\alpha}_t\circ
\int_0^t
\bm{\alpha}_s^{-1}\circ \sqrt{-2(\log \bm{\alpha}_s)'}\circ \d\bm{W}_s.
\label{eq:Ht_decomp}
\end{equation}

Next we identify the law of $\bm{Z}_t$. Since the integrand in \eqref{eq:Ht_decomp}
is deterministic, $\bm{Z}_t$ is a centered Gaussian vector. Moreover, under the standard
assumption that the Brownian motion $\bm{W}_t$ is independent of the initial condition
$\bm{H}_0$, the random vector $\bm{Z}_t$ is independent of $\bm{H}_0$.
Because the diffusion coefficient is diagonal and the coordinates of $\bm{W}_t$ are
independent, the coordinates of $\bm{Z}_t$ are also independent. Thus, we only need to discuss the element-wise variation of $\bm{Z}_t$ to obtain its distribution.

Specifically, for each
$i\in\{1,\dots,d\}$,
\begin{align}
\mathrm{Var}(Z_{t,i})
&=
\alpha_{t,i}^2
\int_0^t
\frac{-2(\log\alpha_{s,i})'}{\alpha_{s,i}^2}\d\bm{s}.
\label{eq:var_pre}
\end{align}
Through simple calculus computation, we have the result
\begin{align}
\mathrm{Var}(Z_{t,i})
&=
\alpha_{t,i}^2
\int_0^t
\frac{d}{ds}\left(\alpha_{s,i}^{-2}\right)\,ds \notag\\
&=
\alpha_{t,i}^2
\left(\alpha_{t,i}^{-2}-\alpha_{0,i}^{-2}\right) \notag\\
&=
\alpha_{t,i}^2
\left(\alpha_{t,i}^{-2}-1\right)
=
1-\alpha_{t,i}^2=\beta_{t,i}^2.
\end{align}

Thus, we know $\bm{Z}_t\sim\mathcal{N}\left(\bm{0},\textrm{diag}(\bm{\beta}_t^2)\right)$.
Hence there exists $\bm{\xi}\sim\mathcal{N}(\mathbf{0},\mathbb{I})$, independent of $\bm{H}_0$,
such that $\bm{Z}_t=\bm{\beta}_t\circ\bm{\xi}$.
Combining this with \eqref{eq:Ht_decomp}, we conclude that
\begin{equation}
\bm{H}_t
\overset{d}{=}
\bm{\alpha}_t\circ \bm{H}_0 + \bm{\beta}_t\circ \bm{\xi},
\end{equation}
where $\overset{d}{=}$ means having the same distribution. Then, the distribution in \eqref{SDE_eq} becomes obvious, which completes the proof.
\section{Proof to Theorem \ref{the:DDIM}}\label{app:proof}
This proof mainly follows that of Theorem 2.8 in \cite{proof}.
Since $\mu$ has compact support, following Lemma 2.6 in \cite{proof}, we know that $\mathcal{D}_t^{\bm{\alpha}}$ is Lipschitz continuous and satisfies the linear growth condition. Hence, ODE \eqref{ODE_inverse} admits a unique global flow. For the initial law specified in Theorem \ref{the:DDIM}, namely $\bm G_{t_0}\sim \phi^{\bm{\alpha}}_{T-t_0}$, we denote by $\psi_t^{\bm{\alpha}}$ the density of $\bm G_t$ whenever it exists.

From the definition of $\rho_{\bm{\beta}}$, we can first verify that
\begin{equation}
{\partial_{t}} \rho_{{\bm \beta}_t} (\bm{x})=(\bm{\beta}_t')^\top\left(\bm{\beta}_t^{-3}\circ\bm{x}^2-\bm{\beta}_t^{-1}\right)\rho_{{\bm \beta}_t}(\bm{x}),
\end{equation}
and
\begin{equation}
\mathrm{div} \left((\bm{\beta}_t^2)'\circ\nabla\rho_{{\bm \beta}_t}(\bm{x})\right)=2(\bm{\beta}_t')^\top\left(\bm{\beta}_t^{-3}\circ\bm{x}^2-\bm{\beta}_t^{-1}\right)\rho_{{\bm \beta}_t}(\bm{x}).
\end{equation}
Thus,
\begin{equation}
{\partial_{t}} \rho_{{\bm \beta}_t} (\bm{x})=\frac{1}{2}\mathrm{div} \left((\bm{\beta}_t^2)'\circ\nabla\rho_{{\bm \beta}_t}(\bm{x})\right).
\end{equation}

Then, we have
\begin{equation}
\begin{aligned}
\partial_t\phi^{\bm{\alpha}}_t(\bm{h})
=&\int_{\mathcal{R}^d}\frac{1}{2}\mathrm{div} \left((\bm{\beta}_t^2)'\circ\nabla\rho_{{\bm \beta}_t}(\bm{x})\right)\mu(\d \bm{\varepsilon})\\
&-\int_{\mathcal{R}^d}(\bm{\alpha}_t'\circ{\bm{\varepsilon}})^\top\nabla \rho_{\bm{\beta}_t}(\bm{h}-\bm{\alpha}_t\circ{\bm{\varepsilon}})\mu(\d \bm{\varepsilon})\\
=&\frac{1}{2}\mathrm{div}\left\{(\bm{\beta}_t^2)'\circ\nabla\phi^{\bm{\alpha}}_t(\bm{h})\right\}\\
&-\mathrm{div}\int_{\mathcal{R}^d}(\bm{\alpha}_t'\circ{\bm{\varepsilon}})\rho_{\bm{\beta}_t}(\bm{h}-\bm{\alpha}_t\circ{\bm{\varepsilon}})\mu(\d \bm{\varepsilon}),
\end{aligned}
\end{equation}
where the operator $\mathrm{div}$ is operated on variable $\bm{h}$.
Noticing that
\begin{equation}
\begin{aligned}
&\nabla\phi^{\bm{\alpha}}_t(\bm{h})\\=&-\int_{\mathcal{R}^d}((\bm{h}-\bm{\alpha}_t\circ{\bm{\varepsilon}})\circ{\bm{\beta_t^{-2}}})\rho_{\bm{\beta}_t}(\bm{h}-\bm{\alpha}_t\circ{\bm{\varepsilon}})\mu(\d \bm{\varepsilon})\\
=&\bm{\alpha}_t\circ\bm{\beta}^{-2}_t\circ\int_{\mathcal{R}^d}{\bm{\varepsilon}}\rho_{\bm{\beta}_t}(\bm{h}-\bm{\alpha}_t\circ{\bm{\varepsilon}})\mu(\d \bm{\varepsilon})-\bm{\beta}^{-2}_t\circ\bm{h}\phi^{\bm{\alpha}}_t(\bm{h}),
\end{aligned}
\end{equation}
we further obtain
\begin{equation}
\begin{aligned}
&\partial_t\phi^{\bm{\alpha}}_t(\bm{h})\\
=&\mathrm{div}\left\{(\bm{\alpha}_t\circ[\log\bm{\beta}_t]'-\bm{\alpha}_t')\circ\int_{\mathcal{R}^d}{\bm{\varepsilon}}\rho_{\bm{\beta}_t}(\bm{h}-\bm{\alpha}_t\circ{\bm{\varepsilon}})\mu(\d \bm{\varepsilon})\right.\\
&\left.-[\log\bm{\beta}_t]'\circ\bm{h}\phi^{\bm{\alpha}}_t(\bm{h})\right\}.
\end{aligned}
\end{equation}
As $\bm{\alpha}_t=\sqrt{1-\bm{\beta}_t^2}$, we have 
\begin{equation}\bm{\alpha}_t'=-\bm{\beta}_t\circ\bm{\alpha}_t^{-1}\circ\bm{\beta}_t'=(\bm{\alpha}_t-\bm{\alpha}^{-1}_t)\circ[\log\bm{\beta}_t]'.\end{equation}
Thus,
\begin{equation}\footnotesize
\begin{aligned}
&\partial_t\phi^{\bm{\alpha}}_t(\bm{h})\\
=&\mathrm{div}\left\{[\log \bm{\beta}_t]'\circ\left[\bm{\alpha}_t^{-1}\circ\int_{\mathcal{R}^d}\bm{\varepsilon}\rho_{\bm{\beta}_t}(\bm{h}-\bm{\alpha}_t\circ \bm{\varepsilon})\mu(\d \bm{\varepsilon})-\bm{h}\phi^{\bm{\alpha}}_t(\bm{h})\right]\right\}\\
=&\mathrm{div}\left\{[\log \bm{\beta}_t]'\circ\left[\bm{\alpha}_t^{-1}\circ\mathbb{E}_{\bm{\varepsilon}\sim\mu}\bm{\varepsilon}\rho_{\bm{\beta}_t}(\bm{h}-\bm{\alpha}_t\circ \bm{\varepsilon})-\bm{h}\phi^{\bm{\alpha}}_t(\bm{h})\right]\right\}\\
=&\mathrm{div}\left\{[\log \bm{\beta}_t]'\circ\phi^{\bm{\alpha}}_t(\bm{h})(\bm{\alpha}_t^{-1}\circ\mathcal{D}_t^{\bm{\alpha}}(\bm{h})-\bm{h})\right\},
\end{aligned}
\end{equation}
Further, by the definition of $\bm b_t^{\bm{\alpha}}(\bm g)$, we have
\begin{equation}\label{phi-b}
    \partial_t \phi^{\bm{\alpha}}_{T-t}(\bm g)
    =
    -\operatorname{div}\!\left(
    \bm b_t^{\bm{\alpha}}(\bm g)\,\phi^{\bm{\alpha}}_{T-t}(\bm g)
    \right).
\end{equation}

On the other hand, for any $f\in C_b^1(\mathcal{R}^d)$, the ODE flow of $\bm G$ satisfies
\begin{equation}
    f(\bm G_t)
    =
    f(\bm G_{t_0})
    +
    \int_{t_0}^t
    \bm b_s^{\bm{\alpha}}(\bm G_s)\cdot \nabla f(\bm G_s)\,\mathrm ds.
\end{equation}
Taking expectations on both sides, and writing the expectation at time $s$ by the density $\psi_s^{\bm{\alpha}}$, we obtain
\begin{equation}
\begin{aligned}
\int_{\mathcal{R}^d} f(\bm g)\psi_t^{\bm{\alpha}}(\bm g)\,\mathrm d\bm g
&=
\int_{\mathcal{R}^d} f(\bm g)\psi_{t_0}^{\bm{\alpha}}(\bm g)\,\mathrm d\bm g \\
&\quad
+
\int_{t_0}^t
\int_{\mathcal{R}^d}
\bm b_s^{\bm{\alpha}}(\bm g)\cdot \nabla f(\bm g)\,
\psi_s^{\bm{\alpha}}(\bm g)\,
\mathrm d\bm g\,\mathrm ds \\
&=
\int_{\mathcal{R}^d} f(\bm g)\psi_{t_0}^{\bm{\alpha}}(\bm g)\,\mathrm d\bm g \\
&\quad
-
\int_{t_0}^t
\int_{\mathcal{R}^d}
f(\bm g)\,
\operatorname{div}\!\left(
\bm b_s^{\bm{\alpha}}(\bm g)\psi_s^{\bm{\alpha}}(\bm g)
\right)
\mathrm d\bm g\,\mathrm ds.
\end{aligned}
\end{equation}
Since this holds for every $f\in C_b^1(\mathcal{R}^d)$, we conclude that, in the weak sense,
\begin{equation}\label{psi-pde}
    \partial_t \psi_t^{\bm{\alpha}}(\bm g)
    =
    -\operatorname{div}\!\left(
    \bm b_t^{\bm{\alpha}}(\bm g)\psi_t^{\bm{\alpha}}(\bm g)
    \right).
\end{equation}
Equivalently,
\begin{equation}\label{psi-int}
    \psi_t^{\bm{\alpha}}(\bm g)
    =
    \psi_{t_0}^{\bm{\alpha}}(\bm g)
    -
    \int_{t_0}^t
    \operatorname{div}\!\left(
    \bm b_s^{\bm{\alpha}}(\bm g)\psi_s^{\bm{\alpha}}(\bm g)
    \right)\,\mathrm ds.
\end{equation}

Define
\[
\varphi_t^{\bm{\alpha}}(\bm g)
:=
\phi^{\bm{\alpha}}_{T-t}(\bm g)-\psi_t^{\bm{\alpha}}(\bm g).
\]
Since $\psi_{t_0}^{\bm{\alpha}}(\bm g)=\phi^{\bm{\alpha}}_{T-t_0}(\bm g)$, by \eqref{phi-b} and \eqref{psi-pde} we obtain
\begin{equation}\label{varphi-pde}
    \partial_t \varphi_t^{\bm{\alpha}}(\bm g)
    =
    -\operatorname{div}\!\left(
    \bm b_t^{\bm{\alpha}}(\bm g)\varphi_t^{\bm{\alpha}}(\bm g)
    \right),
    \qquad
    \varphi_{t_0}^{\bm{\alpha}}(\bm g)=0,
\end{equation}
again in the weak sense. Equivalently,
\begin{equation}\label{varphi-int}
    \varphi_t^{\bm{\alpha}}(\bm g)
    =
    -
    \int_{t_0}^t
    \operatorname{div}\!\left(
    \bm b_s^{\bm{\alpha}}(\bm g)\varphi_s^{\bm{\alpha}}(\bm g)
    \right)\,\mathrm ds.
\end{equation}

Assuming sufficient regularity and decay at infinity to justify differentiation under the integral sign and integration by parts, we have
\begin{equation}
\begin{aligned}
\frac{\mathrm d}{\mathrm dt}\int_{\mathcal{R}^d}\left|\varphi_t^{\bm{\alpha}}(\bm g)\right|^2\,\mathrm d\bm g
&=
2\int_{\mathcal{R}^d}
\varphi_t^{\bm{\alpha}}(\bm g)\,
\partial_t\varphi_t^{\bm{\alpha}}(\bm g)\,
\mathrm d\bm g \\
&=
-2\int_{\mathcal{R}^d}
\varphi_t^{\bm{\alpha}}(\bm g)\,
\operatorname{div}\!\left(
\bm b_t^{\bm{\alpha}}(\bm g)\varphi_t^{\bm{\alpha}}(\bm g)
\right)\,
\mathrm d\bm g \\
&=
2\int_{\mathcal{R}^d}
\varphi_t^{\bm{\alpha}}(\bm g)\,
[\bm b_t^{\bm{\alpha}}(\bm g)]^{\mathrm T}
\nabla\varphi_t^{\bm{\alpha}}(\bm g)\,
\mathrm d\bm g \\
&=
\int_{\mathcal{R}^d}
[\bm b_t^{\bm{\alpha}}(\bm g)]^{\mathrm T}
\nabla\!\left|\varphi_t^{\bm{\alpha}}(\bm g)\right|^2\,
\mathrm d\bm g \\
&=
-\int_{\mathcal{R}^d}
\operatorname{div}\!\left(\bm b_t^{\bm{\alpha}}(\bm g)\right)
\left|\varphi_t^{\bm{\alpha}}(\bm g)\right|^2\,
\mathrm d\bm g.
\end{aligned}
\end{equation}

For every $\varepsilon>0$, the coefficient $[\log\bm\beta_{T-t}]'$ is bounded on $[t_0,T-\varepsilon]$, and since $\mathcal D_t^{\bm\alpha}$ is Lipschitz continuous, there exists a finite constant $C_\varepsilon$ such that
\begin{equation}
\frac{\mathrm d}{\mathrm dt}\int_{\mathcal{R}^d}\left|\varphi_t^{\bm{\alpha}}(\bm g)\right|^2\,\mathrm d\bm g
\le
C_\varepsilon
\int_{\mathcal{R}^d}\left|\varphi_t^{\bm{\alpha}}(\bm g)\right|^2\,\mathrm d\bm g,
\qquad t\in[t_0,T-\varepsilon].
\end{equation}
Since $\varphi_{t_0}^{\bm{\alpha}}(\bm g)=0$, Gr\"onwall's inequality implies that
\[
\varphi_t^{\bm{\alpha}}(\bm g)=0,
\qquad \forall\, t\in[t_0,T-\varepsilon].
\]
Because $\varepsilon>0$ is arbitrary, we conclude that
\[
\varphi_t^{\bm{\alpha}}(\bm g)=0,
\qquad \forall\, t\in[t_0,T).
\]
Hence,
\[
\psi_t^{\bm{\alpha}}(\bm g)=\phi_{T-t}^{\bm{\alpha}}(\bm g),
\qquad \forall\, t\in[t_0,T),
\]
which proves \eqref{eq:DDIM}. The statement at $t=T$ follows by letting $t\uparrow T$.

\end{appendices}
\printbibliography

\end{document}